\documentclass[apj,onecolumn,numberedappendix]{openjournal}
\usepackage{newtxtext,newtxmath,enumitem,multirow,ctable}
\usepackage[T1]{fontenc}
\usepackage{xcolor}
\usepackage[breaklinks,colorlinks,citecolor=blue,urlcolor=blue]{hyperref}

\newcommand{\rthis}[1]{\textcolor{black}{#1}}

\newcommand{\orcidauthor}[3]{\author{\href{http://orcid.org/#1}{#2$^{#3}$}}}

\begin{document}

\title{\vspace{-0.8cm} A Targeted Gamma-Ray Search of Five Prominent Galaxy Merger Systems  with   17 years of Fermi-LAT Data}

\email{* ph22resch11006@iith.ac.in}
\email{** shntn05@gmail.com}

\orcidauthor{0009-0007-1942-8794}{Siddhant Manna}{1 *}
\orcidauthor{0000-0002-0466-3288}{Shantanu Desai}{1 **} 
\affiliation{$^{1}$Department of Physics, IIT Hyderabad Kandi, Telangana 502284, India.}

\begin{abstract}
Galaxy mergers are among the most energetic astrophysical phenomena, driving intense star formation and potentially fueling cosmic ray acceleration, which can produce high energy $\gamma$-ray emission through hadronic processes. We present a targeted search for $\gamma$-ray emission from five prominent galaxy merger systems, NGC~3256, NGC~660, UGC~813/816, UGC~12914/12915, and VV~114 using 16.9 years of Fermi-LAT data in the 1--300~GeV energy range. Employing a binned maximum likelihood analysis, we model the emission with power-law spectra and derive spectral energy distributions (SEDs) to constrain $\gamma$-ray fluxes and spectral indices. Marginal detections are found for NGC~3256 (TS = 15.4, $\sim$3.51$\sigma$) and NGC~660 (TS = 8.16, $\sim$2.39$\sigma$), with photon fluxes of $(7.21 \pm 3.17) \times 10^{-11}$ and $(8.28 \pm 3.56) \times 10^{-11}$ ph cm$^{-2}$ s$^{-1}$, respectively, suggesting merger driven star formation contributes to $\gamma$-ray emission. The remaining systems yield non-detections (TS $< 5$). This is the first targeted study of $\gamma$-ray emission from these aforementioned galaxy merger systems.
\end{abstract}


\maketitle

\section{Introduction}
\label{sec:intro}
Galaxy mergers are one of the most energetic phenomena in the Universe. These  gravitational interactions trigger complex processes that reshape galaxies, funneling gas toward their central regions and fueling intense episodes of star formation, known as nuclear starbursts, as well as potentially enhancing the activity of central supermassive black holes (SMBHs)~\citep{Barnes1992,Hopkins2006}. The resulting extreme environments create ideal conditions for the acceleration of cosmic rays (CRs), high energy particles that interact with interstellar gas and radiation fields through hadronic processes to produce high-energy $\gamma$-ray emission~\citep{Lacki2011,Kashiyama2014}. Such processes include pion decay from proton-proton interactions and inverse Compton scattering of relativistic electrons, which are particularly pronounced in dense, star-forming regions often associated with mergers~\citep{Abdo2010}. Luminous and ultra-luminous infrared galaxies (LIRGs and ULIRGs), frequently observed in late stage mergers, are characterized by their intense infrared emission, which is driven by dust heated by star formation and active galactic nuclei (AGN) activity~\citep{Sanders1996}. These systems have been proposed as significant contributors to the extragalactic $\gamma$-ray background, a diffuse flux of high-energy photons observed across the sky~\citep{Ackermann2012}. However, direct detections of $\gamma$-ray emission from individual galaxy mergers remain scarce, with only a handful of nearby starburst galaxies, such as NGC~253 and M82, providing clear evidence of $\gamma$-ray production~\citep{Abdo2010,Ackermann2012}. These detections serve as critical benchmarks for understanding cosmic ray propagation, energy loss mechanisms, and calorimetry in high density environments, offering insights into the physical processes governing $\gamma$-ray emission in merging systems.

The Fermi Gamma ray Space Telescope, launched in 2008, has revolutionized the study of high energy astrophysical phenomena through its primary instrument, the Large Area Telescope (LAT)~\citep{Atwood2009}. \rthis{Operating over an energy range from 20~MeV to beyond 500~GeV, the LAT offers broad energy coverage with high sensitivity and good angular resolution, particularly above a few hundred~MeV, where it achieves sub-degree localization and peak performance. However, both the sensitivity and angular resolution degrade significantly at lower (tens of MeV) energies compared to the GeV range and above~\footnote{For detailed LAT performance characteristics, see \url{https://www.slac.stanford.edu/exp/glast/groups/canda/lat_Performance.htm}.}.}
These capabilities have enabled all sky surveys that have cataloged thousands of $\gamma$-ray sources~\citep{Abdollahi2020}. The Fourth Fermi Source Catalog Data Release 4 (4FGL-DR4), based on 14 years of LAT observations, represents the deepest survey of the $\gamma$-ray sky to date in the 50 MeV to 1 TeV energy range~\citep{4FGLDR3,Ballet2023}. Despite these advancements, many interacting galaxy systems remain undetected in $\gamma$-rays, and systematic searches targeting galaxy mergers are limited, underscoring the need for dedicated studies to probe their high-energy emission.

Recent multimessenger studies have further highlighted the potential of galaxy mergers as sites of high-energy particle production. For instance, a recent analysis by~\citet{Bouri2025} investigated the hypothesis that galaxy mergers could be sources of high energy neutrinos, which are often produced alongside $\gamma$-rays in hadronic interactions. Using 10 years of IceCube muon track data and galaxy merger catalogs, their analysis found no statistically significant correlation between the galaxy mergers studied and high energy neutrino events. The study set upper limits on the neutrino flux at 100 TeV, suggesting that these mergers do not significantly contribute to the diffuse astrophysical neutrino flux detected by IceCube~\citep{Bouri2025}. This lack of neutrino signal motivates further investigation into the $\gamma$-ray emission from these systems, as the presence or absence of $\gamma$ rays can provide complementary constraints on the underlying particle acceleration mechanisms.

In this work, we conduct a targeted search for $\gamma$-ray emission from a sample of five prominent galaxy mergers, NGC~3256, NGC~660, UGC~813/816, UGC~12914/12915, and VV~114 using 16.9 years of Fermi-LAT data. These systems were selected for their diverse merger stages and intense star-forming activity, making them prime candidates for detecting high-energy emission. They have also been studied in~\citet{Bouri2025} and they found no statistical significance. Initial theoretical predictions for $\gamma$-ray emission from galaxy mergers were made by~\citet{Lisenfeld2010}, who analyzed two of the systems in our sample: UGC~12914/12915 and UGC~813/816. They estimated the hadronic $\gamma$-ray flux resulting from neutral pion ($\pi^0$) decay to be on the order of $\mathcal{O}(10^{-15})~\rm{ph~cm^{-2}~s^{-1}}$, assuming a power-law spectral index of 1.1 normalized at 1~TeV.
This work is a follow-up of our previous work, where we carried out a spatial correlation analysis between a Galaxy Zoo based merger catalog and the fourth Fermi point source catalog (4FGL-DR4)~\citep{Manna25}. 
In Section~\ref{sec:data_analysis} we have laid out our data analysis process, which is common for all the merger systems. \rthis{The target selection and galaxies properties are discussed in Section~\ref{sec:targets}.} In Section~\ref{sec:results} we depict our results for all the mergers.\rthis{In Section~\ref{sec:control} we depict the result of our sample control analysis verifying our results}, \rthis{in Section~\ref{sec:flux_comparison} we compare the theoretical flux with our measured values.} and finally in Section~\ref{sec:conclusions} we present our conclusions.

\section{Data Analysis}
\label{sec:data_analysis}

We performed a targeted search for $\gamma$-ray emission from the galaxy mergers in our sample: NGC~3256, NGC~660, UGC~813/816, UGC~12914/12915, and VV~114, using 16.9~years of Fermi-LAT observations spanning 2008 August 5 to 2025 June 25 (Mission Elapsed Time: 239587201 to 772502405)~\citep{Atwood2009}. \rthis{While our analysis incorporates 16.9~years of data, the Fourth Fermi-LAT Source Catalog Data Release 4 (4FGL-DR4) is based on the first 14~years of observations~\citep{Ballet2023}. Our extended dataset thus allows us to probe fainter or long-term $\gamma$-ray emission beyond the cataloged sources.}  

Data reduction and analysis were carried out using the \texttt{easyfermi} pipeline~\citep{deMenezes2022}, which integrates \texttt{Fermipy}~\citep{Wood2017}, \texttt{Gammapy}~\citep{Donath2023}, \texttt{Astropy}~\citep{Astropy2018}, and \texttt{emcee}~\citep{Foreman2013}. The pipeline automates a binned likelihood analysis by instantiating the \texttt{GTAnalysis} class from \texttt{Fermipy} using a configuration file (\texttt{config.yaml}) generated from user inputs. This tool has previously been applied in searches for $\gamma$-ray emission  in our previous works on magnetars~\citep{Vyaas} and OJ287~\citep{Pasumarti}.  

\rthis{To visualize the spatial distribution of the $\gamma$-ray signal, we generated a test statistic (TS) map for each target system. The TS is defined as}:
\begin{equation}
\mathrm{TS} = -2 \ln \left( \frac{\mathcal{L}_{\max,0}}{\mathcal{L}_{\max,1}} \right),
\end{equation}
\rthis{where $\mathcal{L}_{\max,0}$ is the maximum likelihood for the null hypothesis (background only), and $\mathcal{L}_{\max,1}$ is the maximum likelihood including the source. Under Wilks' theorem~\citep{Wilks1938}, TS follows an asymptotic half  $\chi^2$ distribution in the null case, and the detection significance is approximately given by  $\sqrt{\rm TS}$~\citep{Mattox1996, Manna2024}.} 

\rthis{Each TS map was computed assuming a power-law spectrum with a fixed photon index of 2.0 and covers a $10^\circ \times 10^\circ$ region of interest (ROI) centered on the optical position of the merger. The map color scale represents TS values up to 25, and contours corresponding to $\sqrt{\mathrm{TS}} = 3, 4,$ and $5$ (i.e., TS = 9, 16, and 25) are overlaid in orange, deepskyblue, and lime, respectively. The optical position of the target galaxy is indicated with a cyan star, while the white cross marks the location of the maximum TS if it deviates from the optical center. Nearby 4FGL-DR4 sources within $10^\circ$ of the ROI are indicated by green circles with catalog names labeled.}

\subsection{Event Selection and ROI Setup}
\label{subsec:event_roi}

To ensure high-quality data, we applied standard Fermi-LAT selection cuts (\texttt{DATA\_QUAL > 0} and \texttt{LAT\_CONFIG == 1}) and excluded photons with zenith angles greater than $105^\circ$ to minimize contamination from Earth limb emission, particularly significant at low energies~\citep{Abdo2009}. For each target, we defined a circular region of interest (ROI) of $10^\circ$ radius centered on the galaxy merger coordinates. To account for sources near the ROI boundary, all cataloged sources within an additional $10^\circ$ beyond the ROI radius were included in the model (\texttt{src\_roiwidth = roi\_width + 10}). Energy dispersion corrections were applied (\texttt{edisp = True}), and point-spread function (PSF) convolution was enabled to model the LAT’s energy-dependent angular resolution, which improves significantly above 1~GeV~\citep{Atwood2009}.

\subsection{Background Modeling}
\label{subsec:background}

We employed the \texttt{P8R3\_SOURCE\_V3} instrument response functions (IRFs) for point-source analyses, along with the recommended Galactic diffuse emission model (\texttt{gll\_iem\_v07.fits}) and isotropic background template (\texttt{iso\_P8R3\_SOURCE\_V3\_v1.txt}). \rthis{While identical to those used in 4FGL-DR4~\citep{Ballet2023}, we did not apply additional modifications to the Galactic diffuse component, nor include the Sun and Moon templates. For the purposes of this work, this background configuration is sufficient for modeling the ROI emission.}

All 4FGL-DR4 sources within the ROI were included. Sources with catalog TS~$<16$ had their spectral parameters fixed to 4FGL-DR4 catalog values~\citep{deMenezes2022}; if the likelihood fit did not converge, these sources were temporarily removed and the fit rerun. Sources with TS~$\gtrsim 16$ had their normalizations left free, and particularly bright sources (TS~$\gtrsim 50$) had both normalization and spectral index free during the fit~\citep{deMenezes2022}. This approach ensures that nearby catalog sources are properly accounted for, minimizing systematic biases in the target flux estimation~\citep{deMenezes2022}.

\subsection{Target Source Modeling}
\label{subsec:target_model}

For each target, a test point source was injected at the ROI center and modeled with a power-law spectrum, initialized with a photon index of 2.0~\citep{Ackermann2012}. The injected source was treated independently of existing 4FGL-DR4 sources. During likelihood fitting, both the normalization and photon index were left free to vary. 

\subsection{Iterative Search for Additional Background Sources}
\label{subsec:iterative_bg}

\rthis{Residual maps were generated after fitting the baseline background model to identify any unmodeled excesses. We adopted a detection threshold of $\sqrt{\mathrm{TS}} > 5$ (corresponding to $\sim 5\sigma$) and required a minimum angular separation of $0.5^\circ$ between peaks~\citep{Mattox1996}. Significant residuals were iteratively added as additional point sources with power-law spectra, with both normalization and photon index left free during subsequent likelihood fitting. This process was repeated until no further excesses above threshold remained~\citep{deMenezes2022}. For the galaxy mergers analyzed here, no sources with $\sqrt{\mathrm{TS}} > 5$ were identified. Remaining hotspots in the TS maps correspond to regions with $9 < \mathrm{TS} \leq 16$. These hotspots are quite far from the center of the ROI so their effect is negligible.}

\subsection{Likelihood Analysis and Spectral Energy Distribution}
\label{subsec:likelihood_sed}

The analysis was performed over 1–300~GeV, divided into five logarithmically spaced energy bins for spectral energy distribution (SED) modeling. The lower threshold of 1~GeV mitigates the degraded LAT PSF at lower energies~\citep{Atwood2013}. A binned likelihood analysis was performed with $0.1^\circ$ spatial bins and eight energy bins per decade, using the \texttt{NEWMINUIT} optimizer~\citep{James1975}.  

For each target, the SED was computed using the \texttt{sed()} function in \texttt{Fermipy} as implemented in \texttt{easyfermi}, performing independent likelihood fits in each energy bin while fixing the spectral shape to a power-law. \rthis{For energy bins with TS~$<9$ ($<3\sigma$), we derived 95\% confidence upper limits using the profile-likelihood method, which identifies the flux at which the log-likelihood decreases by 1.35 from its maximum~\citep{Ackermann2012}. This yields frequentist upper limits, sufficient for the scope of this work.}

\section{Target Selection and Galaxy Properties}
\label{sec:targets}

\rthis{The selection of our targets was driven by the primary goal of searching for gamma-ray emission from galaxies undergoing intense star formation episodes triggered by mergers. Such environments are expected to be significant cosmic-ray (CR) reservoirs, where CRs are accelerated in supernova remnants (SNRs) and possibly other sites such as stellar winds and massive star clusters. The subsequent interaction of these CRs with the dense interstellar medium (ISM) should produce detectable high-energy gamma-rays via hadronic processes, primarily through proton-proton collisions leading to neutral pion decay ($\pi^0 \rightarrow \gamma\gamma$). 
Our targets were selected based on the following criteria: (1) they are confirmed, well-studied galaxy mergers with clear evidence of merger-induced activity; (2) they are relatively nearby to maximize the expected gamma-ray flux and angular resolution; (3) they possess substantial molecular gas reservoirs to provide adequate target material for CR interactions. The sample spans different merger stages, from early interactions to late-stage, coalesced remnants, allowing us to probe the evolution of CR populations throughout the merger sequence.
A summary of their key properties is presented in Table~\ref{tab:merger_properties}, and detailed descriptions of each system are provided below.}

\subsection{NGC~3256}
\label{ngc3256}

\rthis{NGC~3256 is one of the most luminous infrared galaxies (LIRGs) in the local Universe, located at a distance of approximately $35$--$39$~Mpc~\citep{Sanders2003,Lira2008}, corresponding to a redshift of $z = 0.0094$. It is classified as a late-stage merger resulting from the collision of two gas-rich disk galaxies of roughly equal mass. The coalescence event is estimated to have occurred approximately $500$~Myr ago~\citep{Sargent1989,Aalto1995,Zepf1999,Lira2002}. The system exhibits a highly disturbed morphology, characterized by extended tidal tails spanning $\sim100$~kpc, stellar bridges, and two distinct nuclei separated by approximately $5\farcs0$ (corresponding to $\sim900$~pc)~\citep{Lira2002,Lira2008,Sakamoto2014}. 
The star formation rate is estimated to be $\text{SFR} \approx 50~M_\odot~\text{yr}^{-1}$~\citep{Sakamoto2014,Michiyama2018}. In another study,~\citet{Michiyama2020} derived a total star formation rate of $49 \pm 2~M_\odot~\text{yr}^{-1}$, with contributions from the nuclear and disk-wide starbursts of approximately $34\%$ and $66\%$, respectively. The high level of starburst activity in the nuclear regions indicates ongoing massive star formation and frequent supernova explosions. The supernova rate has been estimated to be $\nu_{\text{SN}} \approx 0.7~\text{yr}^{-1}$ by~\citet{Lira2008}, while~\citet{Norris1995} derived a rate of about $0.3~\text{yr}^{-1}$ in each nucleus based on radio continuum analysis, suggesting that both nuclear regions are prolific sites of massive star death and cosmic-ray acceleration. Such a high supernova frequency implies efficient cosmic-ray production and a potentially detectable level of hadronic gamma-ray emission if the dense interstellar medium provides sufficient target material. The angular size of NGC~3256 is approximately $3.8^{\prime} \times 2.1^{\prime}$.
NGC~3256 is extremely rich in molecular gas, with a total molecular gas mass of $M(\text{H}_2)_{\text{total}} \approx 5 \times 10^{10}~M_\odot$~\citep{Sakamoto2014,Michiyama2018}. Based on CO observations,~\citet{Sargent1989} reported a total gas mass of $3 \times 10^{10}~M_\odot$. This places NGC~3256 among the most gas-rich mergers known, comparable to only a few of the brightest systems in the nearby Universe. The exceptionally high gas surface density in the nuclear regions provides an efficient target for cosmic-ray interactions, making NGC~3256 one of the most promising nearby candidates for detecting hadronic gamma-ray emission.}

\subsection{NGC~660}

\rthis{NGC~660 is a rare and peculiar polar-ring galaxy located at a distance of approximately $13$--$14$~Mpc~\citep{Vandriel1995,Springbob2009}, corresponding to a redshift of $z = 0.0028$~\citep{Yu2022}, making it one of the nearest examples of this unusual class of galaxies. It was likely formed through a galaxy collision or accretion event approximately one billion years ago , resulting in a unique morphology where a late-type lenticular or edge-on spiral disk is surrounded by a gas-rich outer ring~\citep{Whitmore1990,Aalto1995,Vandriel1995,Salter2024}.
It is classified as a LINER (Low-Ionization Nuclear Emission-line Region) galaxy~\citep{Salter2024}. This places NGC~660 in the luminous infrared galaxy category, though at the lower end compared to the most extreme starbursts. The total star formation rate is estimated to be $\text{SFR} \approx (14.3  \pm 0.19)$ $M_\odot~\text{yr}^{-1}$~\citep{Nazri2021}. While NGC~660 shows evidence of intense massive star formation in the nucleus , its overall star formation efficiency is moderate compared to major merger-driven starbursts. 
It has been mapped extensively in both atomic (HI) and molecular (CO) gas , revealing a substantial reservoir of interstellar material. The total gas mass of NGC~660 is approximately $8.5\times10^{9}~M_\odot$, consisting of about $5.4\times10^{9}~M_\odot$ in neutral hydrogen (H\,\textsc{i}) and $3.7\times10^{9}~M_\odot$ in molecular gas (H$_2$), with roughly $75\%$ of the H\,\textsc{i} gas located in the polar ring.~\citep{Combes1992,Vandriel1995}.
The angular size of NGC~660 on the sky is approximately $4\farcm57 \times 1\farcm69$. NGC~660's unique morphology, ongoing star formation in both the disk and polar ring, and complex nuclear activity make it an interesting target for gamma-ray studies, offering insights into cosmic-ray acceleration and transport in a non-standard galaxy environment shaped by past interactions.}

\subsection{The Taffy Galaxies: UGC~12914/12915}
\rthis{The Taffy system consists of the interacting spiral galaxy pair UGC~12914 and UGC~12915, located at a distance of approximately 60$-$62~Mpc, corresponding to a redshift of $z \approx 0.0145$~\citep{Appleton2022}. This system represents one of the most spectacular examples of a nearly head-on galaxy collision, earning its nickname from the radio continuum bridge whose brightness contours stretch between the two galaxies like strands of taffy~\citep{Condon1993}. The collision is estimated to have occurred approximately $20$--$25$~Myr ago based on spectral steepening of synchrotron emission due to radiative losses, dynamical modeling, and mass estimates from HI rotation curves~\citep{Condon1993,Braine2003,Zhu2007}.
During the encounter, the stellar disks of both galaxies largely passed through each other with minimal disruption due to the collisionless nature of stars, while the gaseous components experienced direct cloud-cloud collisions and were stripped from the disks~\citep{Condon1993,Zhu2007}. This resulted in the formation of a massive interstellar medium (ISM) bridge between the two galaxies, containing significant amounts of both atomic and molecular gas. The individual galaxies maintain relatively undisturbed stellar disks with ongoing star formation, while the bridge region exhibits dramatically different properties. Both galaxies are classified as LINER galaxies, indicating the presence of weakly ionized gas in their nuclear regions.
Using H$\alpha$ observations,~\citet{Bushouse1987} derived a total star formation rate of approximately $1.4~M_\odot~\text{yr}^{-1}$ for the Taffy system. Specifically, the measured H$\alpha$ luminosities of $L(\text{H}\alpha) = 1.17 \times 10^{41}$ and $6.31 \times 10^{40}~\text{erg s}^{-1}$ correspond to SFRs of $0.93$ and $0.50~M_\odot~\text{yr}^{-1}$ for UGC~12914 and UGC~12915, respectively, using the calibration of~\citet{Kennicutt1998}. Infrared and mid-infrared analyses yield a wide range of estimates:~\citet{Jarrett1999} reported a low global SFR of $<6~M_\odot~\text{yr}^{-1}$ adopting the conversion factor from~\citet{Roussel2001}, whereas the total infrared luminosity of $L_{\text{IR}} = 7\times10^{10}~L_\odot$ implies a higher SFR of $\sim12~M_\odot~\text{yr}^{-1}$ following~\citet{Kennicutt1998}. However, the reliability of mid-infrared PAH-based SFR estimates in Taffy~I remains uncertain, as small dust grains traced at these wavelengths may have been destroyed by the strong shocks induced by the large-scale galactic collision~\citep{Braine2003}. More recent analyses~\citep{Komugi2012} combining infrared components suggest total SFRs of $\sim7.3$ and $\sim14.9~M_\odot~\text{yr}^{-1}$ for UGC~12914 and UGC~12915, respectively, corresponding to a combined rate of $\sim22~M_\odot~\text{yr}^{-1}$—an order of magnitude higher than that derived from the H$\alpha$ observations.
The most remarkable feature of the Taffy system is its massive gas bridge. Early HI observations revealed approximately $5 \times 10^9~M_\odot$ of atomic hydrogen in the bridge region~\citep{Condon1993}. Subsequent CO observations detected substantial molecular gas throughout the bridge, with initial estimates suggesting $M(\text{H}_2)_{\text{bridge}} \approx 1 \times 10^{10}~M_\odot$~\citep{Braine2003}. The total molecular gas mass in both galactic disks is approximately $M(\text{H}_2)_{\text{disks}} \approx (1$--$2) \times 10^{10}~M_\odot$, bringing the system total to $M(\text{H}_2)_{\text{total}} \approx (1.4$--$2.5) \times 10^{10}~M_\odot$~\citep{Zhu2007,Gao2003,Vollmer2012}.
The Taffy galaxies provide an exceptional laboratory for studying cosmic-ray acceleration, confinement, and transport in a post-collision environment. The presence of significant amounts of molecular gas in the bridge, combined with the demonstrated presence of cosmic rays (evidenced by strong radio synchrotron emission) and magnetic fields, makes this system a compelling target for gamma-ray observations.} 

\subsection{The Taffy II Galaxies: UGC~813/816}

\rthis{The Taffy II system consists of the interacting spiral galaxy pair UGC~813 and UGC~816, representing only the second known example of a nearly head-on spiral-spiral collision after the discovery of the original Taffy galaxies (UGC~12914/12915)~\citep{Condon2002}. Located at a distance of approximately $75-78$~Mpc~\citep{Braine2004}, corresponding to a redshift of $z \approx 0.0145$~\citep{Condon2002,Braine2004}, this system exhibits remarkable morphological and physical properties resulting from the interpenetration of two gas-rich disk galaxies. Like the original Taffy galaxies, UGC~813/816 are joined by a prominent synchrotron radio bridge whose brightness contours resemble stretching bands of taffy, giving the system its nickname~\citep{Condon2002}. The collision is estimated to have occurred approximately $40-50$~Myr ago based on analysis of radio continuum spectral indices, synchrotron aging, and dynamical modeling~\citep{Condon2002,Lisenfeld2008}. 
Based on CO observations, the expected star formation rate in the bridge is about $2~M_\odot~\mathrm{yr^{-1}}$, whereas H$\alpha$ observations indicate a significantly lower value of only $0.1$--$0.4~M_\odot~\mathrm{yr^{-1}}$~\citep{Braine2004}. Using the far-infrared luminosity ($L_{\mathrm{FIR}} \approx 1.5\times10^{10}~h^{2}~L_\odot$) of UGC~813/816, the current star formation rate is estimated to be $\sim1.4~h^{2}~M_\odot~\mathrm{yr^{-1}}$ for massive stars with $M > 5~M_\odot$~\citep{Condon2002}. Despite the enormous amounts of molecular gas in the bridge, the star formation efficiency appears remarkably low in these collisionally-formed molecular clouds~\citep{Braine2004}. This suppressed star formation may result from the expansion of pre-stellar cores during cloud collisions, which retards subsequent gravitational collapse and star formation~\citep{Braine2004}.
A significant fraction of the system's atomic hydrogen lies between the disks in the bridge region. The total H\,\textsc{i} gas mass is of the order $\sim2\times10^{10}~M_\odot$~\citep{Bushouse1987}. Although the H\,\textsc{i} profiles are too distorted for a direct mass measurement, assuming a typical ratio of total-to-hydrogen mass of $M_{\mathrm{T}}/M_{\mathrm{H}} \approx 50$~\citep{Haynes1984}, the total mass of the UGC~813/816 system is estimated to be $M_{\mathrm{T}} \approx 5\times10^{11}~h^{2}~M_\odot$.}

\subsection{VV~114}

\rthis{VV~114 is a luminous infrared galaxy system located at a distance of approximately $78$--$82$~Mpc~\citep{Yun1994,Knop1994,Rich2023}, corresponding to a redshift of $z \approx 0.020$, representing a gas-rich merger in an early to intermediate stage. The system consists of two merging spiral galaxies with a wide separation of approximately $6$~kpc between the optical nuclei, indicating that VV~114 has not yet reached the final coalescence phase~\citep{Yun1994,Kashiyama2014}.
High-resolution CO observations reveal approximately $5.1 \times 10^{10}~M_\odot$ of molecular hydrogen distributed in a $5.9 \times 3.1$~kpc ($15'' \times 8''$) bar-like structure with two prominent tidal tails each extending $4$--$6$~kpc~\citep{Yun1994}. The star formation rate in VV~114 has been estimated approximately as $1~M_\odot~\mathrm{yr^{-1}}$~\citep{Murphy2011,Rich2023}. Similarly ~\citet{Ho2007} gives a SFR in the range of $\sim1.5$--$2.5~M_\odot~\mathrm{yr^{-1}}$. These results are consistent with those derived by~\citet{Song2022} from radio observations.
The system's large molecular gas reservoir, ongoing merger-driven star formation, and early merger stage make VV~114 an interesting target for gamma-ray observations, allowing comparison with more evolved merger systems to study the evolution of cosmic-ray populations throughout the merger sequence.}

\begin{table*}[t]
\centering
\caption{\rthis{Physical properties of the analyzed merger systems.}}
\label{tab:merger_properties}
\begin{tabular}{lccccc}
\hline
\hline
\textbf{System} & \textbf{Distance} & \textbf{Redshift} & \textbf{Merger age} & \textbf{SFR} & \textbf{$M(\mathrm{H}_2)$} \\
 & (Mpc) &  & (Myr) & ($M_\odot\,\mathrm{yr^{-1}}$) & ($M_\odot$) \\
\hline
NGC~3256 & 35--39 & 0.0094 & $\sim500$ & $\sim50$ & $\sim(3$--$5)\times10^{10}$ \\
NGC~660 & 13--14 & 0.0028 & $\sim1000$ & $\sim14$ & $3.7\times10^{9}$ \\
UGC~12914/12915 (Taffy~I) & 60--62 & 0.0145 & 20--25 & $\sim22$ & $(1.4$--$2.5)\times10^{10}$ \\
UGC~813/816 (Taffy~II) & 75--78 & 0.0145 & 40--50 & $\sim1$--$2$ & $\sim10^{10}$ \\
VV~114 & 78--82 & 0.020 & Early/intermediate stage & $\sim1$--$2.5$ & $\sim5.1\times10^{10}$ \\
\hline
\end{tabular}
\end{table*}

\section{Results}
\label{sec:results}
We conducted a comprehensive $\gamma$-ray analysis of five galaxy merger systems—NGC~3256, NGC~660, UGC~813/816, UGC~12914/12915, and VV~114 using 16.9 years of Fermi-LAT data in the 1--300~GeV energy range. The key properties of the analyzed sources are summarized in Table~\ref{tab:gamma_summary}, which lists the source coordinates, TS, photon flux, energy flux, and spectral indices derived from a power law model. For the merger systems with TS < 9, we showcased their 95\% confidence level upper limit values for photon flux and energy flux. 

\begin{table}[htbp]
\centering
\caption{Summary of gamma-ray properties for the detected galaxy merger systems in the 1--300~GeV energy range. For the last three systems as TS < 9, we have depicted 95\% confidence level upper limits for photon and energy flux.}
\label{tab:gamma_summary}
\begin{tabular}{lcccccc}
\hline
\textbf{Galaxy} & \textbf{RA} & \textbf{Dec} & \textbf{TS} & \textbf{Flux} & \textbf{Energy Flux} & \textbf{Spectral Index} \\
 & (deg) & (deg) &  & ($10^{-11}$ ph cm$^{-2}$ s$^{-1}$) & ($10^{-7}$ MeV cm$^{-2}$ s$^{-1}$) & $\Gamma$ \\
\hline
NGC~3256 & 156.96 & -43.90 & 15.40 & $7.21 \pm 3.17$ & $3.77 \pm 1.50$ & $2.05 \pm 0.27$ \\
NGC~660  & 25.76 & 13.65 & 8.16 & $8.28 \pm 3.56$ & $1.81 \pm 0.77$ & $2.83 \pm 0.54$ \\
UGC~12914/12915     & 0.41 & 23.48 & 1.98 & $ < 10.4$ & $ < 1.86$ & $3.25 \pm 0.79$ \\
UGC~813/816  & 19.07  & 46.74 & 2.26 & $ < 5.37$ & $ < 3.62$ & $1.92 \pm 0.60$ \\
VV~114    & 16.95  & -17.51 & 0.02 & $ < 2.47$ & $ < 1.32$ & $2.04 \pm 2.07$ \\
\hline
\end{tabular}
\end{table}

Since the positions of the target sources are known a priori, the likelihood fits involve only two free parameters: the flux normalization and the spectral index. Consequently, the statistical significance of each detection was estimated assuming that the TS follows a $\chi^2$ distribution with two degrees of freedom under the null hypothesis.  

Among the five merger systems analyzed, only NGC~3256 and NGC~660 exhibit marginal broadband $\gamma$-ray signals, with TS values of 15.40 ($p$-value = $2.26 \times 10^{-4}$, $\sim3.51\sigma$) and 8.16 ($p$-value = $8.45 \times 10^{-3}$, $\sim2.39\sigma$), respectively for two degrees of freedom, which correspond to the spectral index and normalization in our case. These low-significance detections are consistent with expectations for galaxy mergers hosting intense star formation and potential cosmic-ray acceleration~\citep{Ackermann2012}.  

The remaining systems—UGC~813/816, UGC~12914/12915, and VV~114 do not show statistically significant emission, with TS values below 9.0, indicating no measurable $\gamma$-ray emission. For comparison, theoretical flux predictions based on hadronic interactions in the shocked regions of the mergers~\citep{Lisenfeld2010} are on the order of $\mathcal{O}(10^{-15})~\mathrm{ph~cm^{-2}~s^{-1}}$ for UGC~12914/12915 and UGC~813/816. Our derived 95\% confidence upper limits are $1.04 \times 10^{-10}$ and $5.37 \times 10^{-11}~\mathrm{ph~cm^{-2}~s^{-1}}$ for UGC~12914/12915 and UGC~813/816, respectively, in the 1--300~GeV band. These limits lie approximately 4–5 orders of magnitude above the theoretical expectations, indicating that the predicted hadronic emission remains well below the current sensitivity of the \textit{Fermi}-LAT.  

We now discuss the results for each system in detail in the following subsections.

\subsection{NGC~3256}
\label{ngc3256:results}

\begin{figure}[ht]
    \centering
    \includegraphics[width=0.65\textwidth]{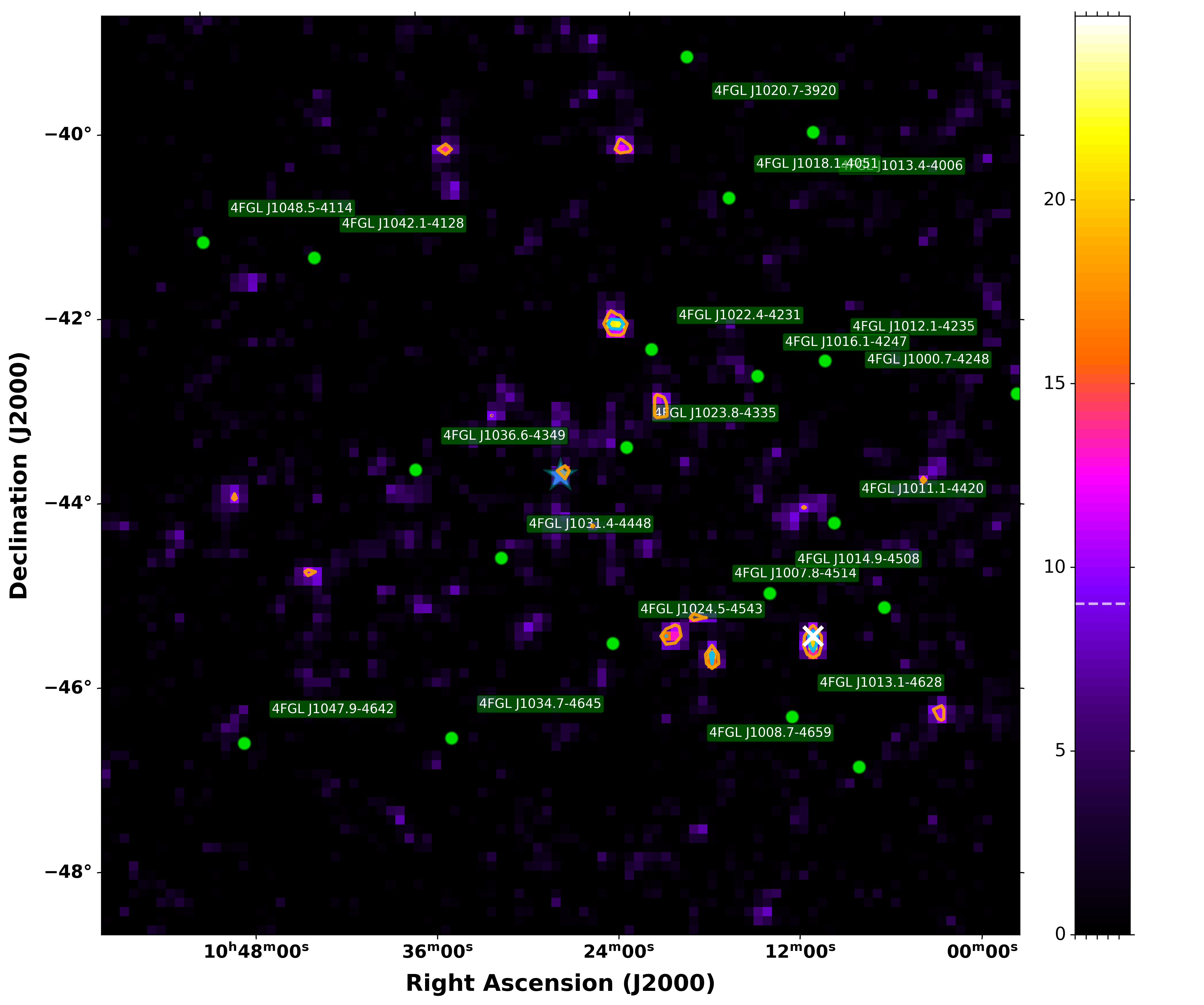}
    \caption{\rthis{TS map of the NGC~3256 in the 1--300~GeV range. The color scale indicates the TS intensity  revealing a localized excess near the target position with TS of 15.40. Contours represent $\sqrt{\mathrm{TS}} = 3$ (orange), 4 (deepskyblue), and 5 (lime). The cyan star marks the optical position of NGC~3256, while the white cross indicates the location of the maximum TS pixel. Nearby 4FGL-DR4 catalog sources are shown as green circles with source names annotated.}}
    \label{fig:Figure1}
\end{figure}

Analysis of 16.9 years of Fermi-LAT observations reveals marginal $\gamma$-ray emission from a point like source spatially coincident with NGC~3256, with a TS value of 15.4, corresponding to a detection significance of approximately $3.51\sigma$. The emission is well described by a power-law spectrum, yielding a photon index of $\Gamma = 2.05 \pm 0.27$, a photon flux of $(7.21 \pm 3.17) \times 10^{-11}~\mathrm{ph~cm^{-2}~s^{-1}}$, and an energy flux of $(3.77 \pm 1.50) \times 10^{-7}~\mathrm{MeV~cm^{-2}~s^{-1}}$ in the 1--300~GeV range. These values are consistent with expectations for starburst galaxies, where $\gamma$-ray emission is primarily attributed to hadronic processes, such as pion decay from cosmic-ray interactions with dense interstellar gas~\citep{Ackermann2012}. Figure~\ref{fig:Figure1} presents the TS map for NGC~3256.

\begin{figure}[ht]
    \centering
    \includegraphics[width=0.65\textwidth]{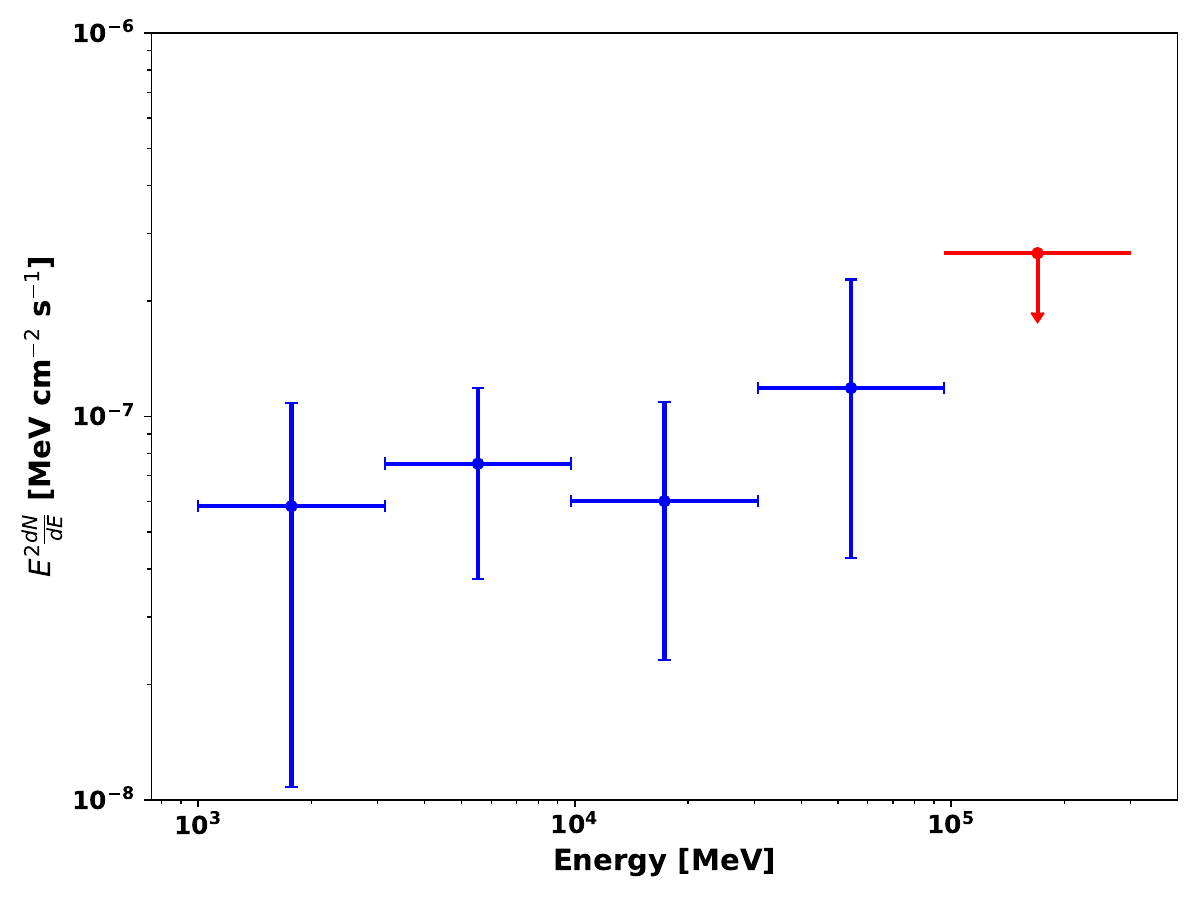}
    \caption{SED of NGC~3256 in the 1--300~GeV range, showing marginal gamma-ray emission. The first four energy bins represent detections with TS > 9, while the last bin shows an upper limit with TS < 9. TS values obtained in the first four bins are 9.39 (2.61$\sigma$), 13.12 (3.19$\sigma$) , 11.73 (2.98$\sigma$) and 13.48 (3.24$\sigma$) respectively.}
    \label{fig:Figure2}
\end{figure}

The spectral energy distribution (SED), shown in Figure~\ref{fig:Figure2}, indicates weak but consistent emission across the 1--300~GeV range. Significant TS values ($>$9, $\sim 3\sigma$) are obtained in the first four of five logarithmically spaced energy bins: 9.39, 13.12, 11.73, and 13.48. The fifth bin yields TS = 7.83 and is therefore treated as an upper limit. The relatively flat photon index suggests a hard spectrum, comparable to other starburst galaxies such as NGC~253 and M82, where pion decay processes dominate~\citep{Abdo2010}. The marginal detection indicates that NGC~3256 is a faint $\gamma$-ray emitter, potentially limited by Fermi-LAT sensitivity. The nearest 4FGL-DR4 catalog source, 4FGL~J1023.8$-$4335, lies $46.8'$ from the target position.

\rthis{The background modeling follows the approach described in Section~\ref{subsec:background}. This treatment ensures an accurate and robust background for estimating the $\gamma$-ray flux of NGC~3256. We note that some nearby 4FGL-DR4 sources could be variable, potentially affecting the flux estimation. However, a dedicated variability analysis is beyond the scope of this work. The impact of source variability is expected to be minor within the ROI and energy range considered. Future studies will assess the influence of variability on the derived fluxes, providing a more comprehensive evaluation of potential systematic effects from neighboring sources.}

\begin{figure}[ht]
\centering
\includegraphics[width=0.6\linewidth]{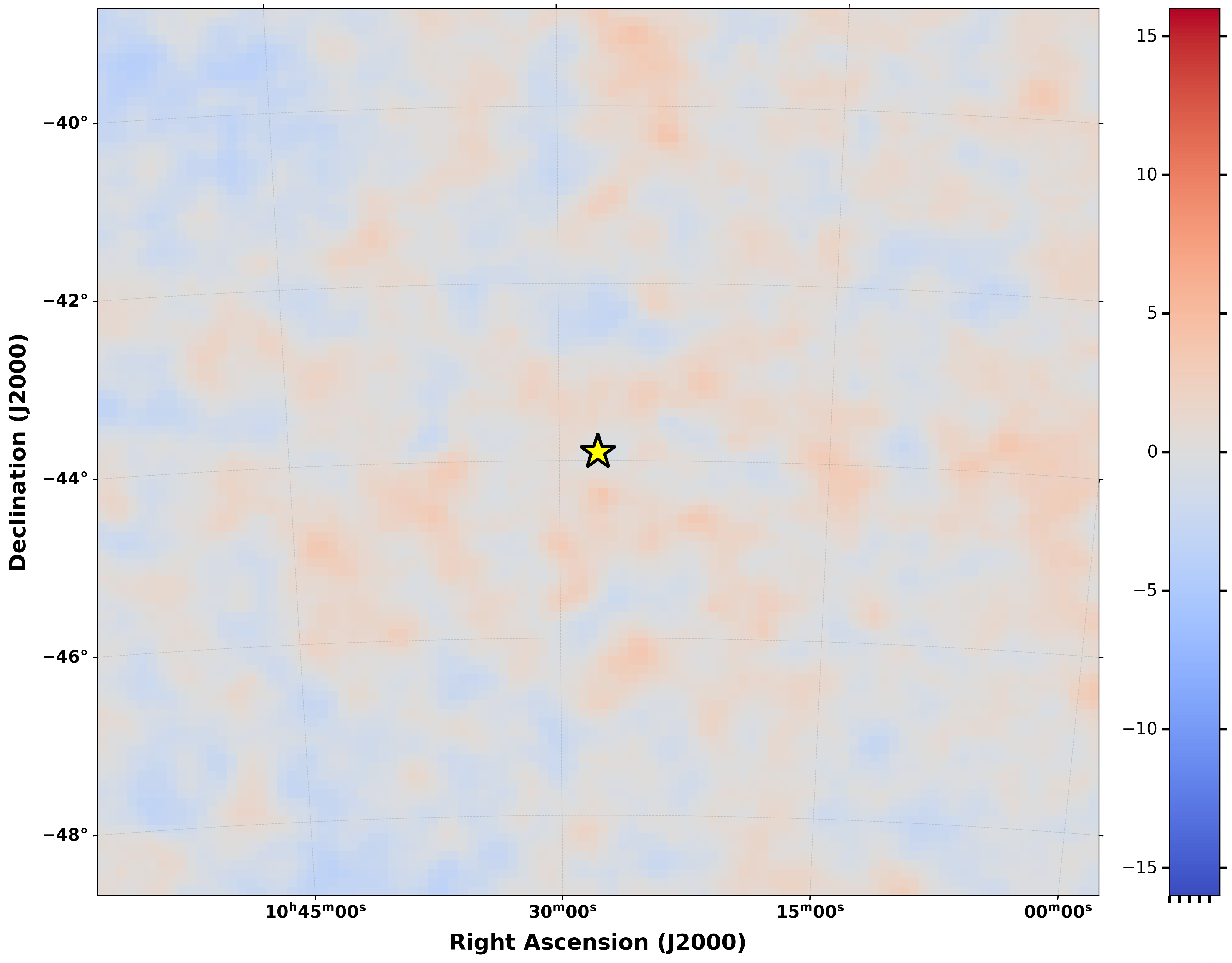}
\caption{\rthis{Residual map of the NGC~3256 field in the 1--300~GeV range. The color bar indicates the TS values ranging from $-16$ to $+16$. The yellow star marks the optical position of NGC~3256.  No significant residual excesses are detected, confirming that the background and source models adequately describe the observed emission.}}
\label{fig:Figure3}
\end{figure}

\rthis{To assess the presence of unmodeled emission, we generated a residual map for NGC~3256. The residual map spans TS values from $-4.21$ to $3.79$, with a mean of $-0.00$ and a standard deviation of 1.17. No pixels exceed $\pm 3\sigma$, indicating that there is no significant unmodeled emission in the ROI. The residual TS map, presented in Figure~\ref{fig:Figure3}, confirms that the observed $\gamma$-ray excess is consistent with the modeled source and background, supporting the robustness of the marginal detection.}

\rthis{Among our merger sample, NGC~3256 exhibits the most significant $\gamma$-ray excess. This galaxy also hosts an X-ray source, 2XMM~J102751.2$-$435413, identified in the XMM-Newton catalog~\citep{Watson2009} and listed in SIMBAD as spatially coincident with NGC~3256. The presence of both X-ray and $\gamma$-ray emission suggests elevated high-energy activity. While this may simply reflect the intense starburst processes in the system, a physical connection between the X-ray and $\gamma$-ray emitting regions cannot be ruled out; a dedicated multiwavelength study would be required to explore this further.}

\subsection{NGC~660}
\label{ngc660}

\begin{figure}[ht]
    \centering
    \includegraphics[width=0.65\textwidth]{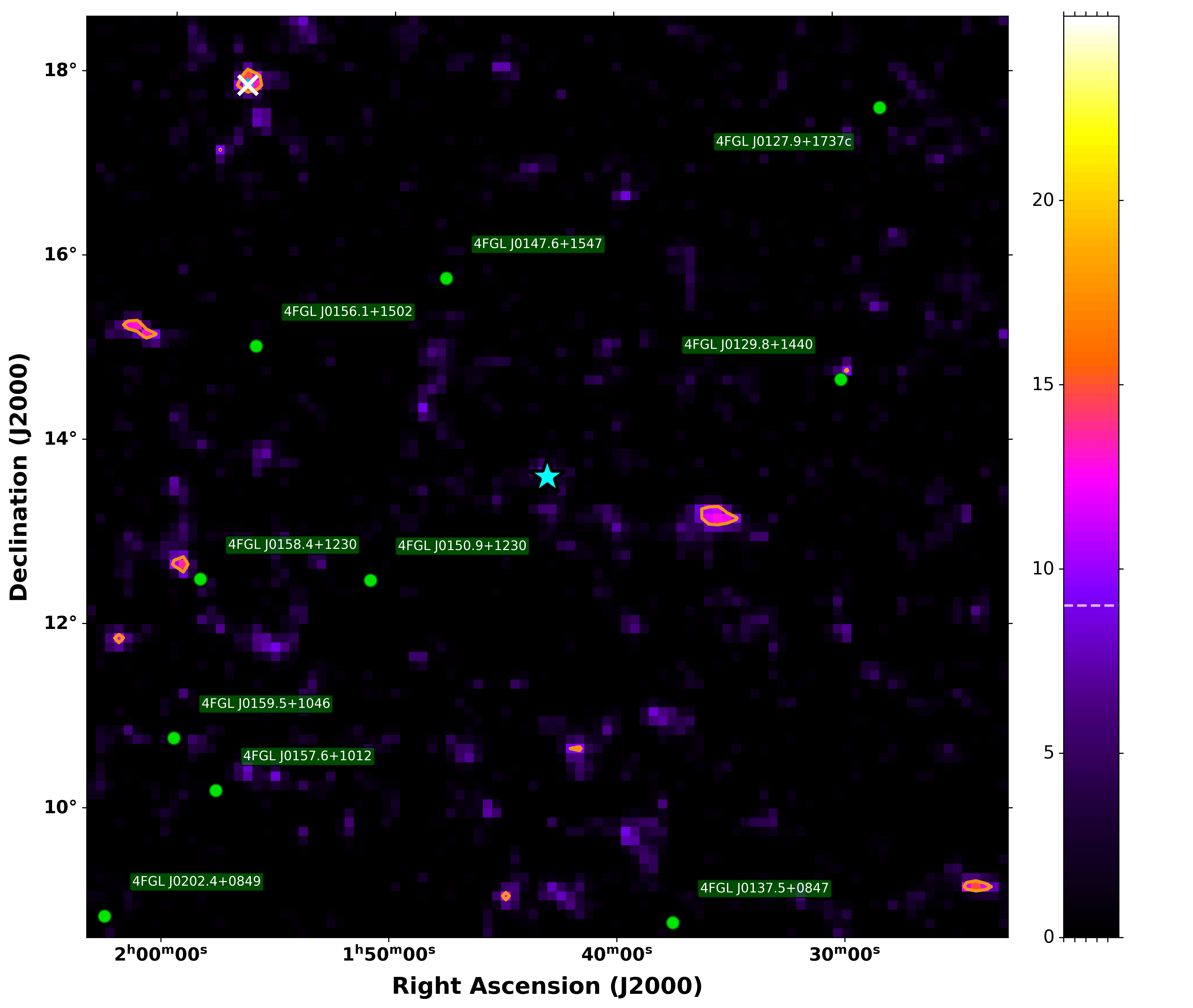}
    \caption{TS map of NGC~660 in the 1--300~GeV range. A mild excess with TS of 8.16 is observed near the target position. The plotting conventions are the same as in Figure~\ref{fig:Figure1}.}
    \label{fig:Figure4}
\end{figure}

Our analysis of 16.9 years of Fermi-LAT observations reveals marginal $\gamma$-ray emission from a point like source spatially coincident with NGC~660, with a TS value of 8.16, corresponding to a detection significance of approximately $2.39\sigma$. The emission is modeled with a power-law spectrum, yielding a photon index of $\Gamma = 2.83 \pm 0.54$, a photon flux of $(8.28 \pm 3.56) \times 10^{-11}~\mathrm{ph~cm^{-2}~s^{-1}}$, and an energy flux of $(1.81 \pm 0.77) \times 10^{-7}~\mathrm{MeV~cm^{-2}~s^{-1}}$ in the 1--300~GeV energy range. Figure~\ref{fig:Figure4} shows the TS map, which indicates a weak $\gamma$-ray excess at the position of NGC~660, consistent with the marginal detection from the likelihood analysis.

\begin{figure}[ht]
    \centering
    \includegraphics[width=0.65\textwidth]{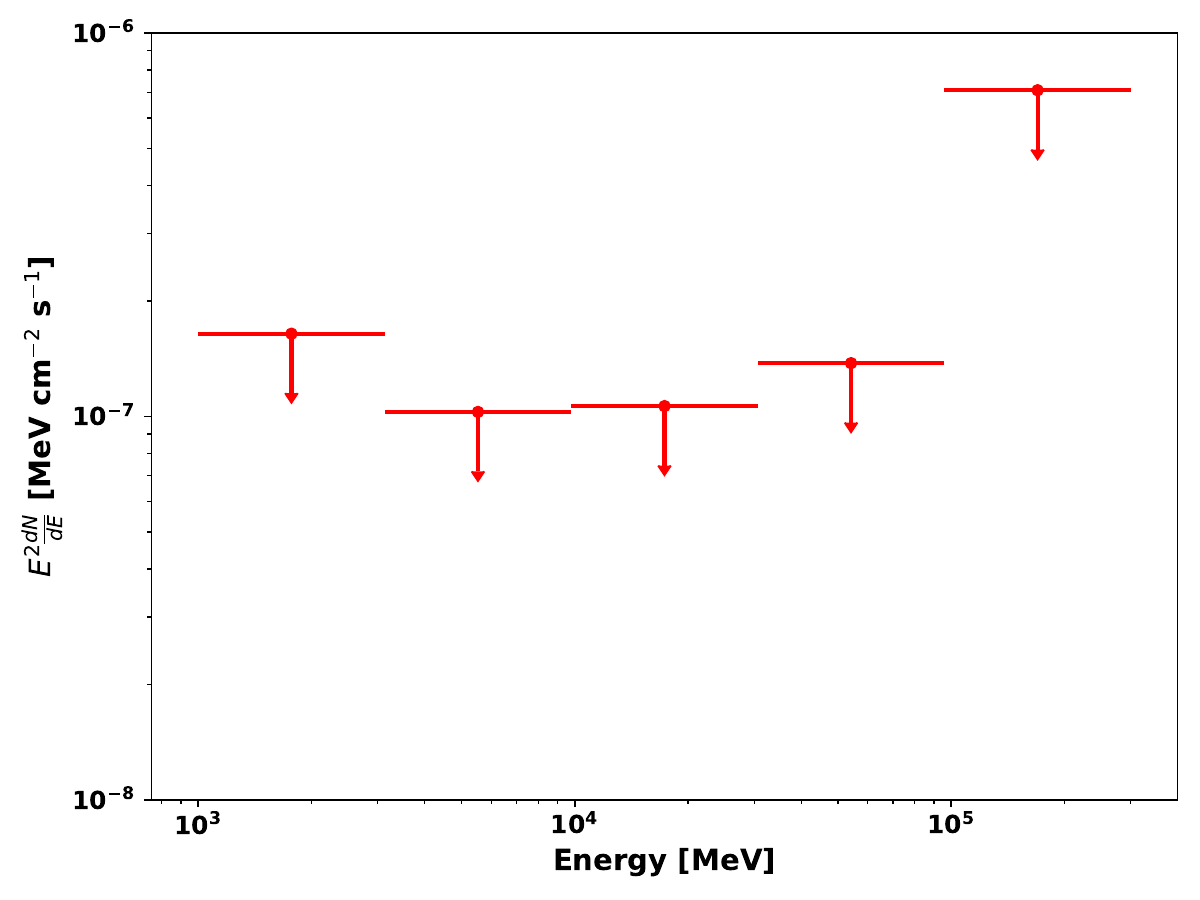}
    \caption{SED of NGC~660 in the 1--300~GeV range, showing upper limits in all five energy bins.}
    \label{fig:Figure5}
\end{figure}

The SED, shown in Figure~\ref{fig:Figure5}, exhibits no statistically significant detection across the five logarithmically spaced energy bins, with TS values of 4.05, 2.66, 0.51, 0, and 0.77, respectively. Consequently, only 95\% confidence upper limits are reported for the flux across the full energy range. The absence of significant emission in individual bins suggests that any $\gamma$-ray signal is faint and broadly distributed over the energy range.

A search of the Fermi-LAT Fourth Source Catalog (4FGL-DR4)~\citep{Ballet2023} confirms that no previously reported $\gamma$-ray sources lie within $1^\circ$ of NGC~660, supporting the interpretation of the observed excess as a marginal signal from the target galaxy.

\rthis{As with NGC~3256, the background was modeled using the standard Pass~8 SOURCE-class diffuse templates and all 4FGL-DR4 sources within the ROI, as described in detail in Section~\ref{subsec:background}. Some neighboring catalog sources may exhibit variability, which could influence the flux estimate; however, a detailed variability study was not conducted in this work. The expected effect on the derived flux for NGC~660 is minor, and future analyses will examine potential systematic effects from variable background sources in more detail.}

\begin{figure}[ht]
\centering
\includegraphics[width=0.6\linewidth]{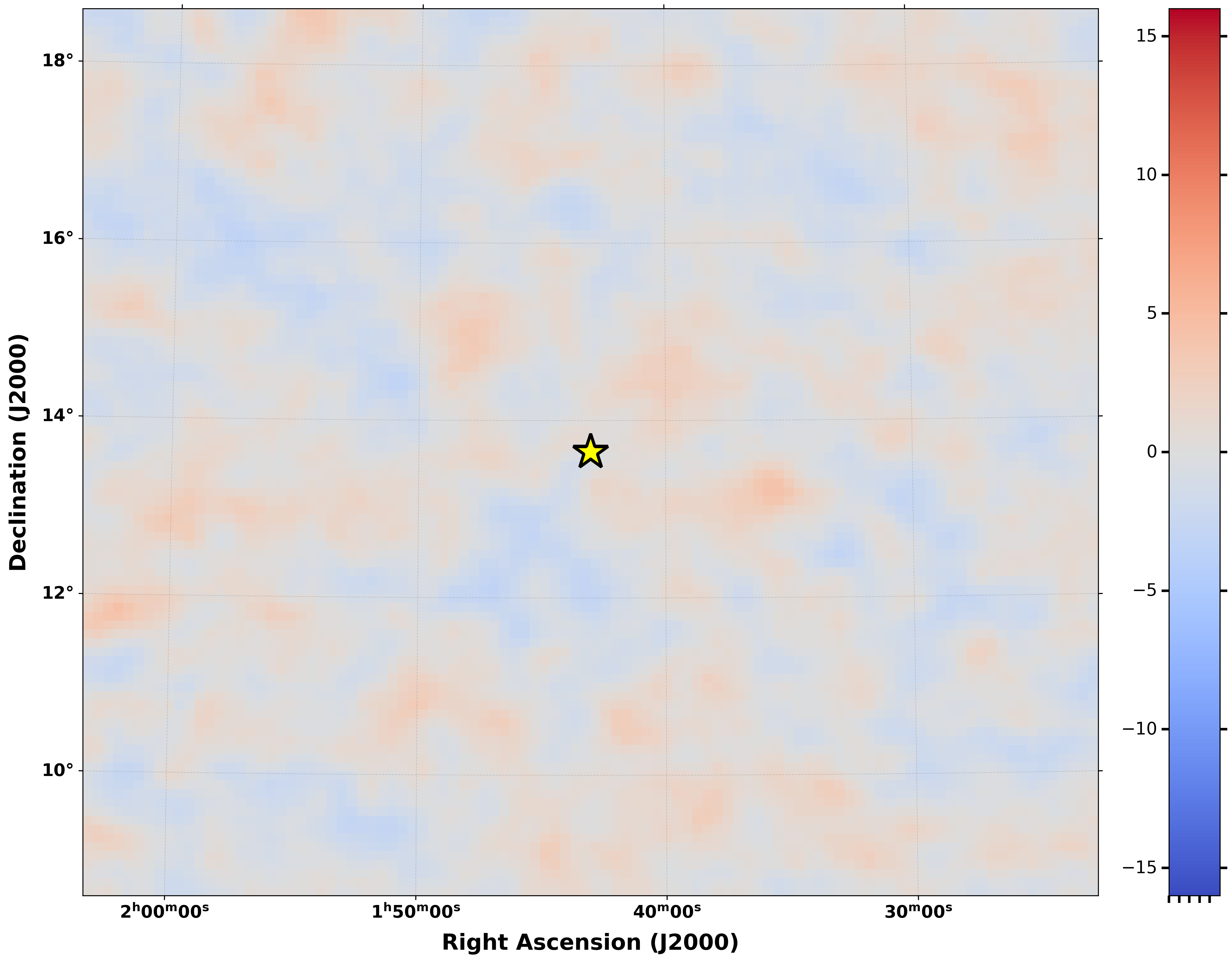}
\caption{\rthis{Residual map of the NGC~660 field in the 1--300~GeV range. The plotting conventions are the same as in Figure~\ref{fig:Figure3}.}}
\label{fig:Figure6}
\end{figure}

\rthis{To further assess the presence of unmodeled emission, we generated a residual map centered on the target coordinates. The TS values in the residual map range from $-3.24$ to $4.40$, with a mean of $-0.02$ and a standard deviation of $1.10$. No pixels exceed $\pm 3\sigma$, indicating the absence of significant residual emission (see Figure~\ref{fig:Figure6}). This residual analysis supports the conclusion that the marginal TS value arises from faint emission consistent with background fluctuations rather than any unmodeled source.}

\subsection{UGC~813/816}
\label{sec:ugc813/6}

\begin{figure}[ht]
    \centering
    \includegraphics[width=0.65\textwidth]{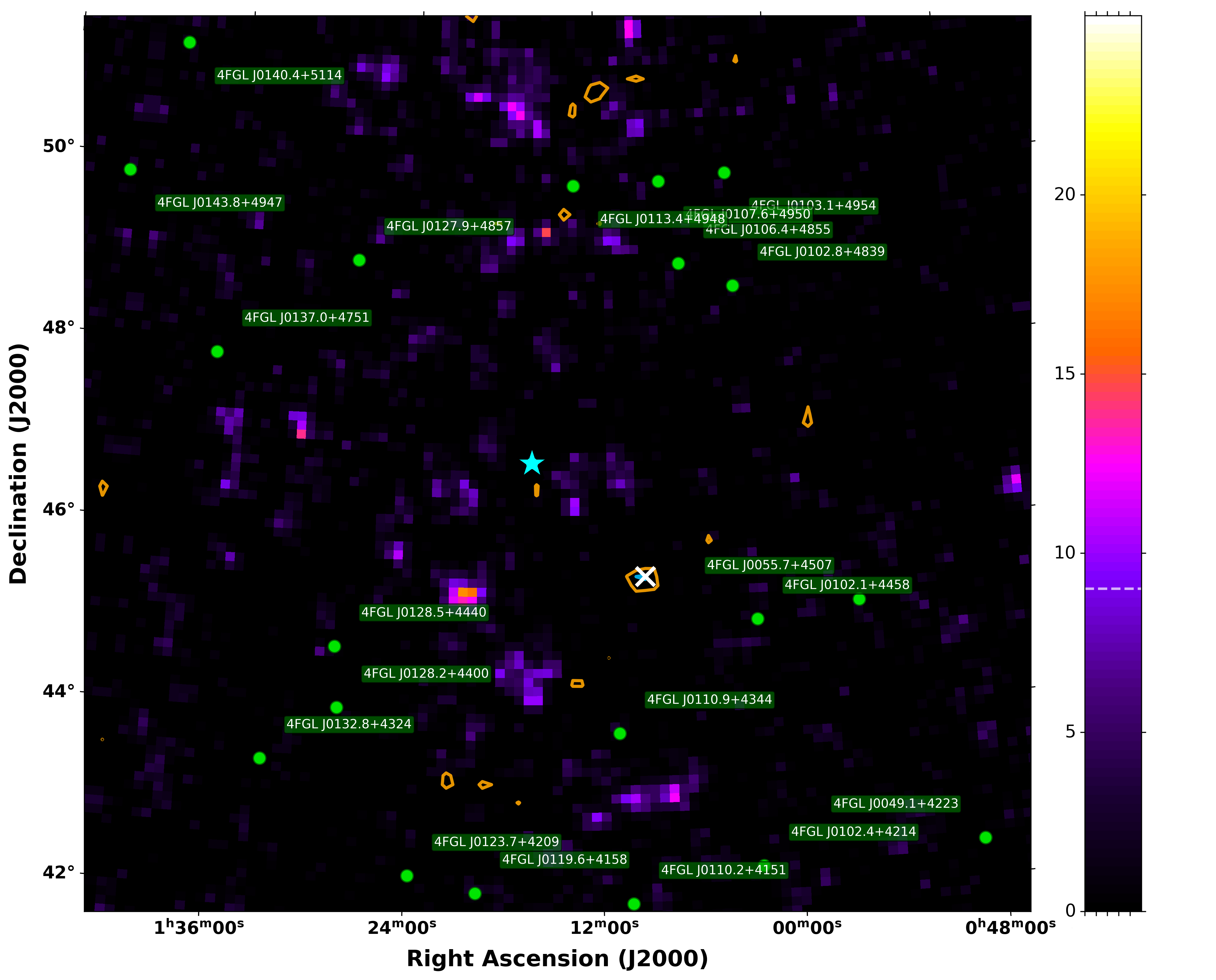}
    \caption{TS map of UGC~813/816 in the 1--300~GeV range. No significant excess is observed in the vicinity, with TS of 2.26. The plotting conventions are the same as in Figure~\ref{fig:Figure1}.}
    \label{fig:Figure7}
\end{figure}

Our analysis of 16.9 years of Fermi-LAT observations yields no statistically significant $\gamma$-ray emission from the UGC~813/816 system, with a TS value of 2.26, corresponding to a detection significance of approximately $0.99\sigma$. The 95\% confidence upper limits are $5.37 \times 10^{-11}~\mathrm{ph~cm^{-2}~s^{-1}}$ for the photon flux and $3.62 \times 10^{-7}~\mathrm{MeV~cm^{-2}~s^{-1}}$ for the energy flux. Figure~\ref{fig:Figure7} shows the TS map, which does not exhibit any significant excess near UGC~813/816, corroborating the non-detection inferred from the likelihood analysis.

\begin{figure}[ht]
    \centering
    \includegraphics[width=0.65\textwidth]{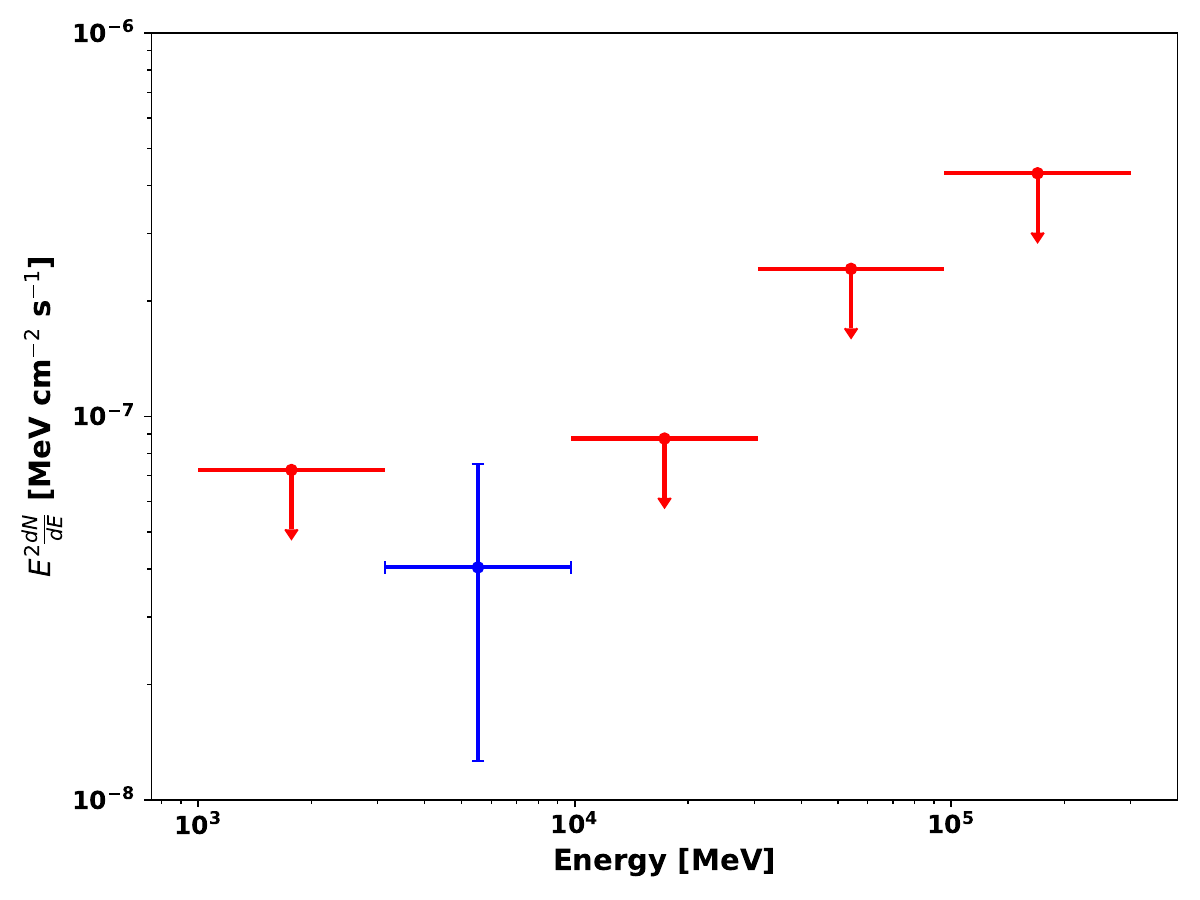}
    \caption{SED of UGC~813/816 in the 1--300~GeV range, showing a detection in the second energy bin with TS of 10.07 (2.72$\sigma$) and upper limits in all other bins.}
    \label{fig:Figure8}
\end{figure}

The SED, displayed in Figure~\ref{fig:Figure8}, shows a marginal TS of 10.07 in the second energy bin, while all other bins remain below the detection threshold (TS = 7.82, 8.25, 7.87, and 7.83) and are therefore reported as upper limits. A search of the \textit{Fermi}-LAT 4FGL-DR4 catalog~\citep{Ballet2023} confirms that no previously reported $\gamma$-ray sources are located within $1^\circ$ of the UGC~813/816 system.

\begin{figure}[ht]
\centering
\includegraphics[width=0.6\linewidth]{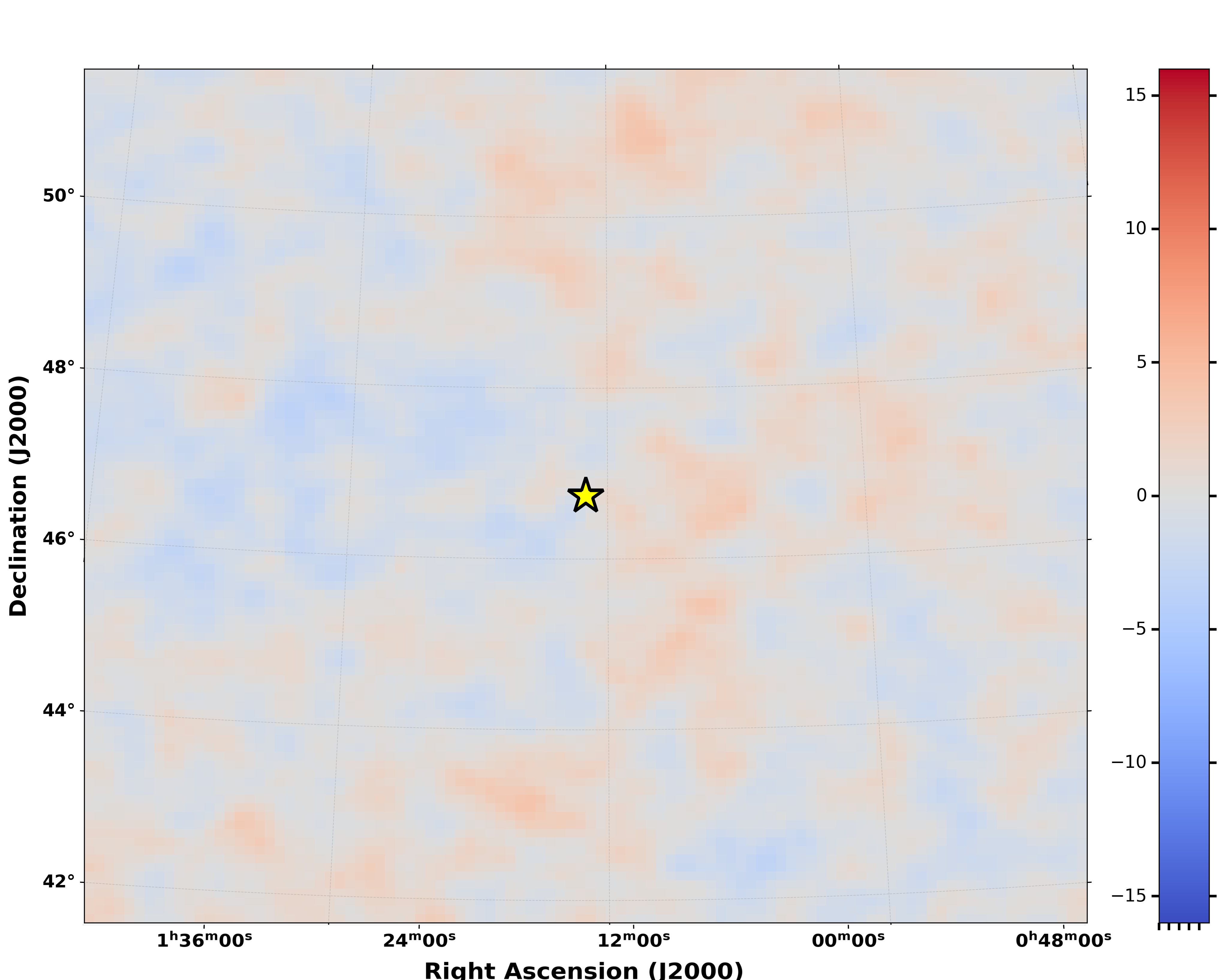}
\caption{\rthis{Residual map of the UGC~813/816 field in the 1--300~GeV range. The plotting conventions are the same as in Figure~\ref{fig:Figure3}.}}
\label{fig:Figure9}
\end{figure}

\rthis{To evaluate potential unmodeled emission, we generated a residual map centered on the UGC~813/816 position. The residual TS values range from $-3.55$ to $4.08$, with a mean of $-0.00$ and a standard deviation of 1.14. No pixels exceed $\pm3\sigma$, indicating that the model adequately describes the observed counts within the region of interest. The absence of significant residuals implies that neither overfitting nor unmodeled $\gamma$-ray structures are present. The residual TS map, shown in Figure~\ref{fig:Figure9}, thus supports the robustness of the background and source modeling and confirms that the observed emission is consistent with statistical fluctuations.}

\subsection{UGC~12914/12915}
\label{sec:ugc12914/5}

\begin{figure}[ht]
    \centering
    \includegraphics[width=0.65\textwidth]{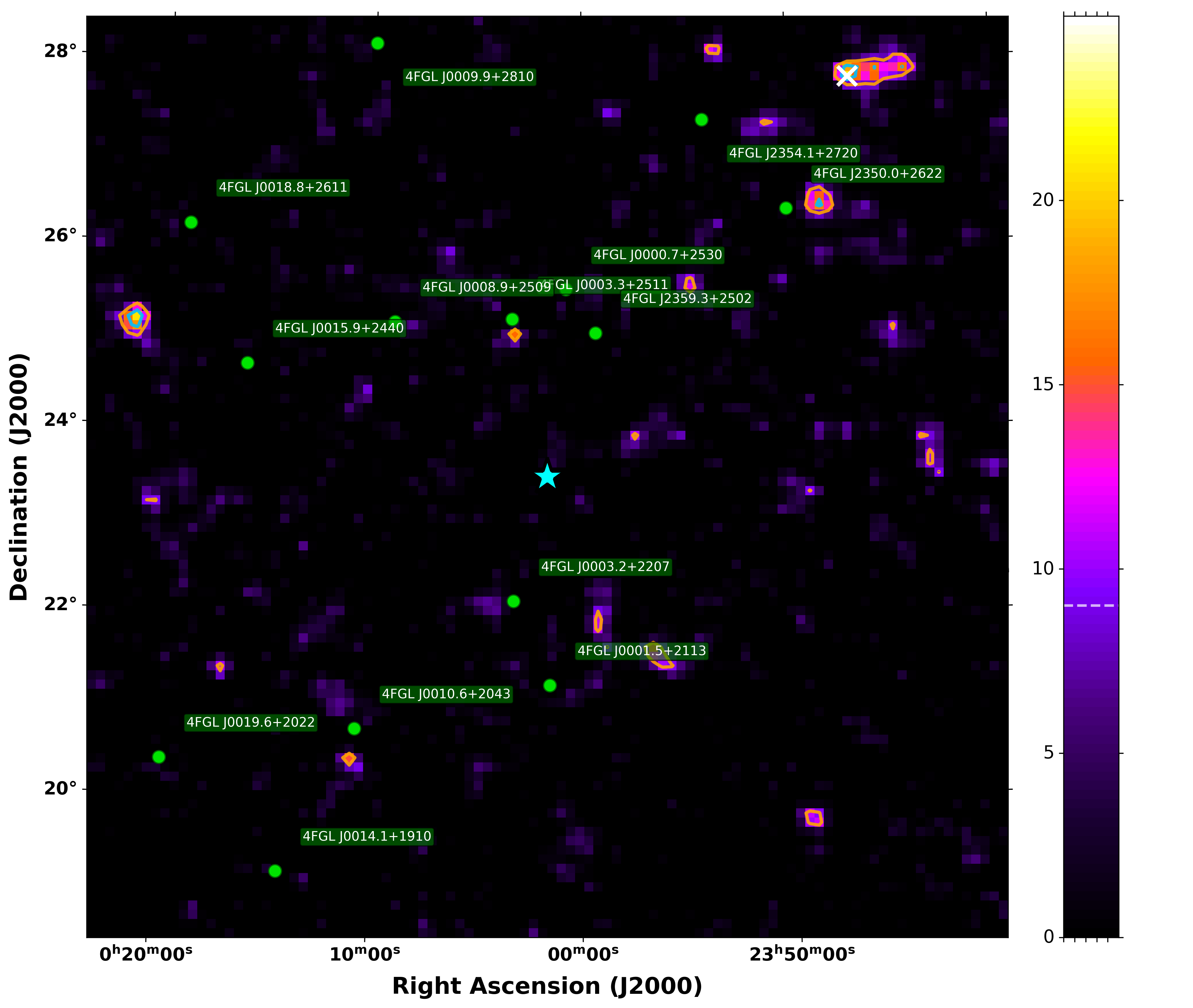}
    \caption{TS map of UGC~12914/12915 in the 1--300~GeV range. No prominent excess is detected, consistent with the TS value of 1.98 from the likelihood analysis. The plotting conventions are the same as in Figure~\ref{fig:Figure1}.}
    \label{fig:Figure10}
\end{figure}

Analysis of 16.9 years of Fermi-LAT observations reveals no statistically significant $\gamma$-ray emission from the UGC~12914/12915 system, with a TS value of 1.98, corresponding to a detection significance of approximately $0.89\sigma$ as shown in~\ref{fig:Figure10}. The 95\% confidence upper limits are $1.04 \times 10^{-10}~\mathrm{ph~cm^{-2}~s^{-1}}$ for the photon flux and $1.86 \times 10^{-7}~\mathrm{MeV~cm^{-2}~s^{-1}}$ for the energy flux.

\begin{figure}[ht]
    \centering
    \includegraphics[width=0.65\textwidth]{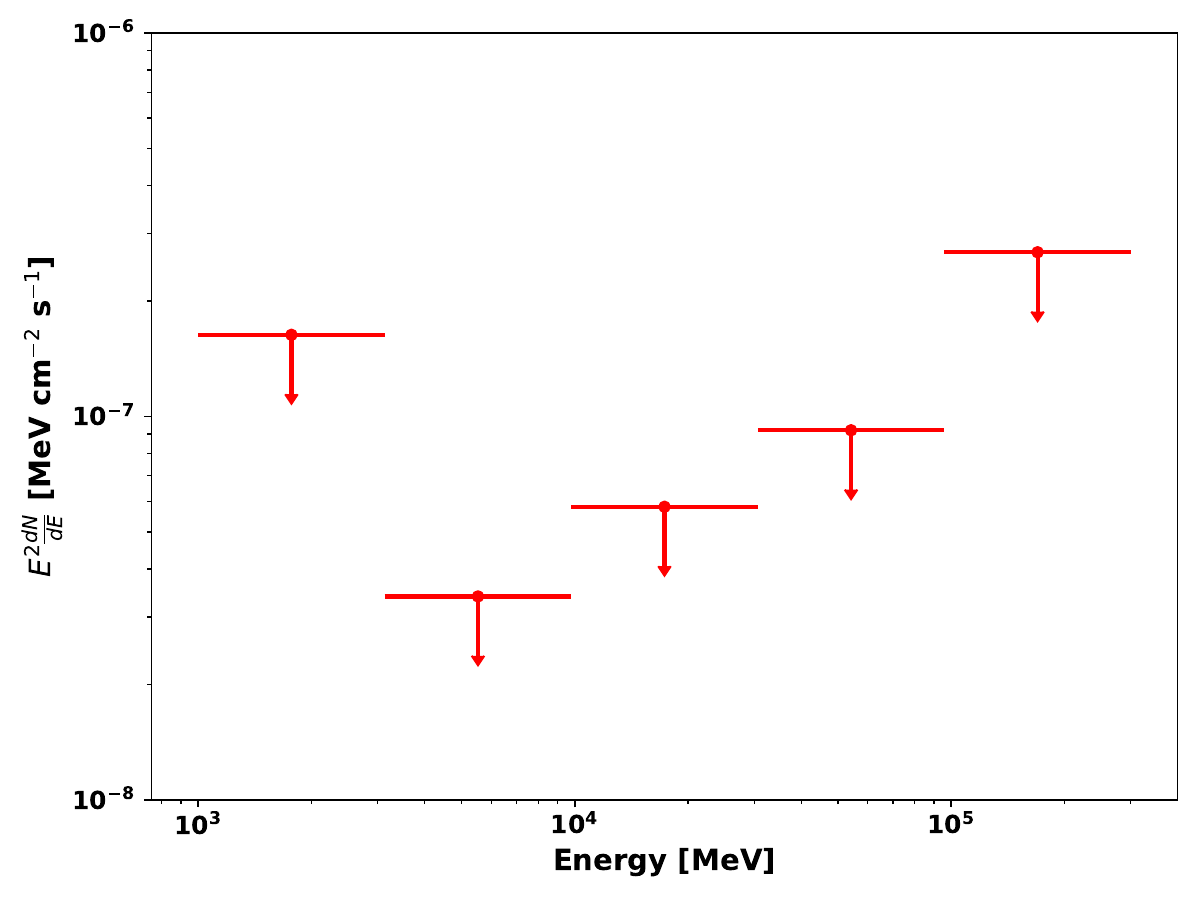}
    \caption{SED of UGC~12914/12915 in the 1--300~GeV range, showing upper limits in all five energy bins.}
    \label{fig:Figure11}
\end{figure}

The SED, shown in Figure~\ref{fig:Figure11}, indicates no significant detection in any of the five logarithmically spaced energy bins, with TS values of 4.71, 0, 0, 0, and 0. Consequently, all bins are reported as upper limits. A search of the Fermi-LAT 4FGL catalog~\citep{Ballet2023} confirms that no previously known $\gamma$-ray sources lie within $1^\circ$ of the UGC~12914/12915 system.

\begin{figure}[ht]
\centering
\includegraphics[width=0.6\linewidth]{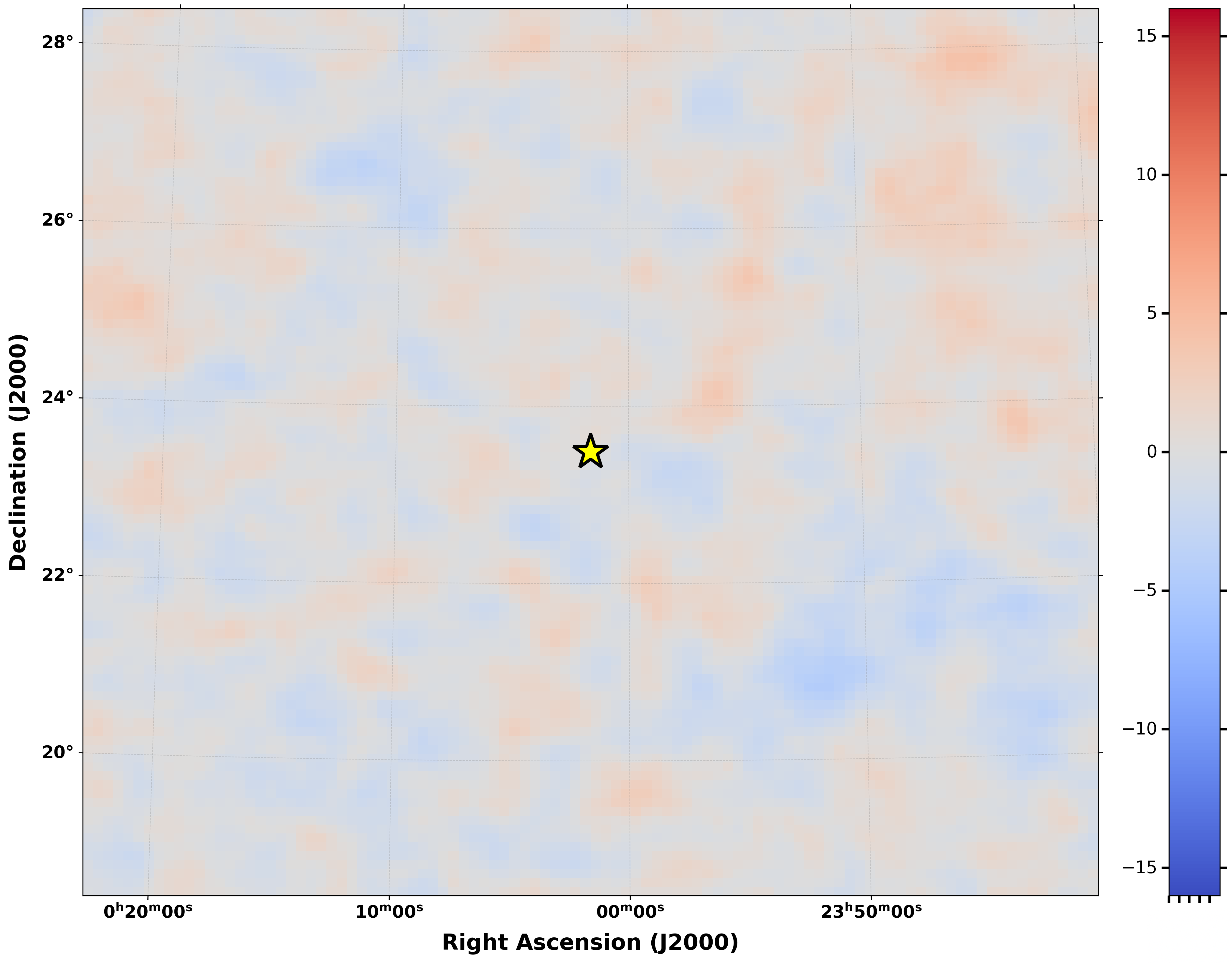}
\caption{\rthis{Residual map of the UGC~12914/12915 field in the 1--300~GeV range. The plotting conventions are the same as in Figure~\ref{fig:Figure3}.}}
\label{fig:Figure12}
\end{figure}

\rthis{To further assess the presence of unmodeled emission, we generated a residual map centered on the target coordinates. The TS values in the map span from $-4.45$ to $4.29$, with a mean of $-0.04$ and a standard deviation of $1.11$, consistent with statistical expectations for background fluctuations. No pixels exceed $\pm 3\sigma$, indicating the absence of significant residual emission. The residual TS map, presented in Figure~\ref{fig:Figure12}, supports the conclusion that there is no detectable $\gamma$-ray signal from the UGC~12914/12915 system.}

\subsection{VV~114}
\label{vv114}

\begin{figure}[ht]
    \centering
    \includegraphics[width=0.65\textwidth]{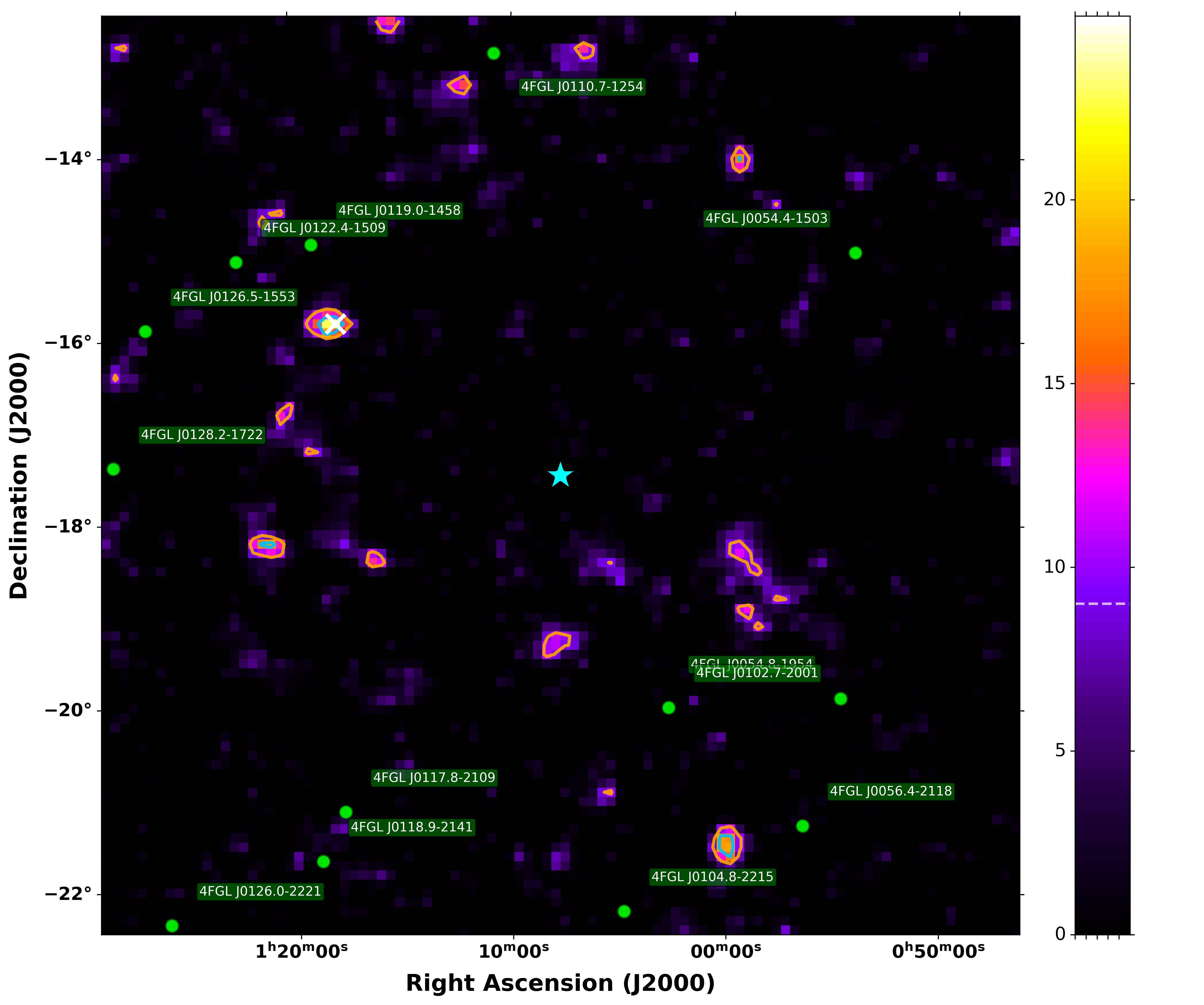}
    \caption{TS map of VV~114 in the 1--300~GeV range. No significant $\gamma$-ray excess is visible with TS of 0.02. The plotting conventions are the same as in Figure~\ref{fig:Figure1}.} 
    \label{fig:vv114_ts_map}
    \label{fig:Figure13}
\end{figure}

Analysis of 16.9 years of Fermi-LAT observations reveals no statistically significant $\gamma$-ray emission from VV~114, with a TS value of 0.02, corresponding to a detection significance of $\sim0.01\sigma$. The 95\% confidence upper limits are $2.47 \times 10^{-11}~\mathrm{ph~cm^{-2}~s^{-1}}$ for the photon flux and $1.32 \times 10^{-7}~\mathrm{MeV~cm^{-2}~s^{-1}}$ for the energy flux. As shown in Figure~\ref{fig:Figure13}, no significant $\gamma$-ray excess is present around the VV~114 system.

\begin{figure}[ht]
    \centering
    \includegraphics[width=0.65\textwidth]{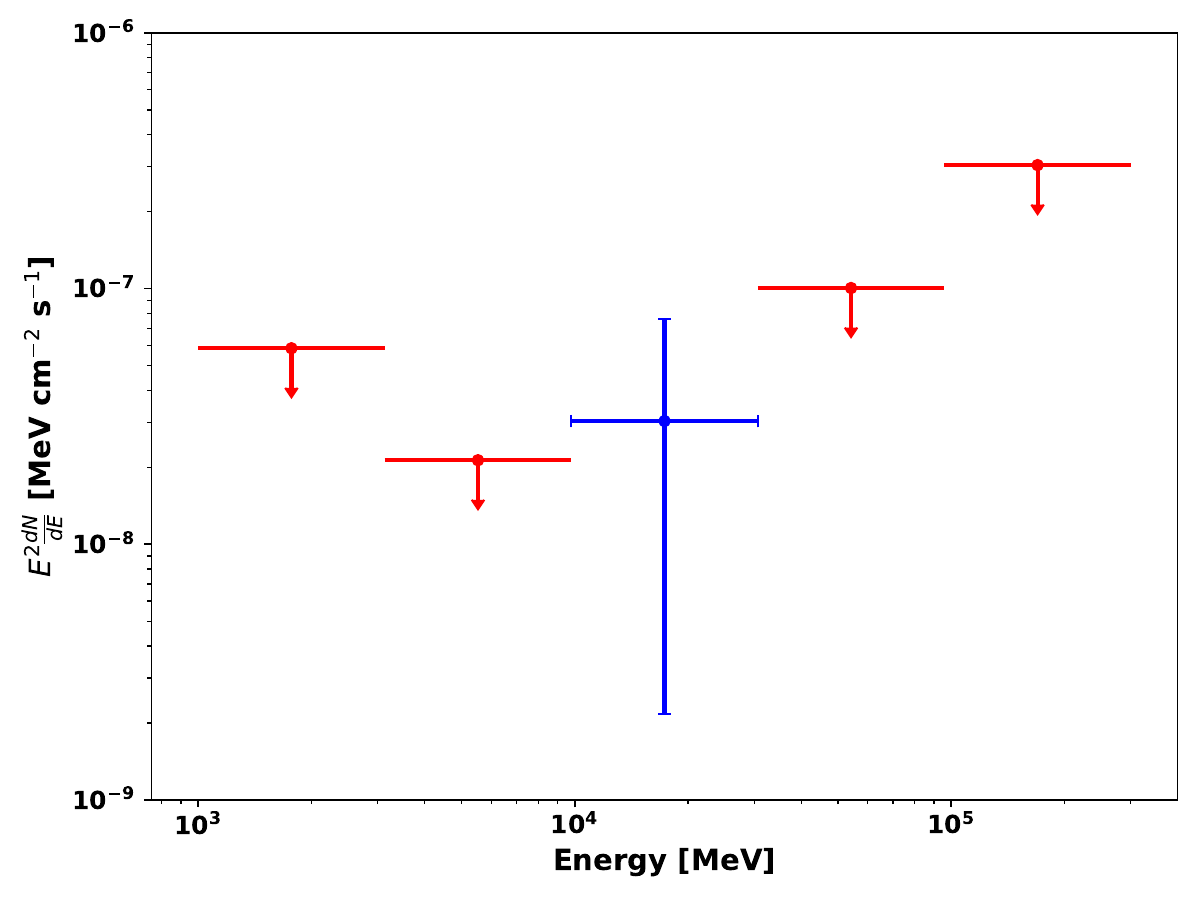}
    \caption{SED of VV~114 in the 1--300~GeV range, showing a detection in the third energy bin and upper limits in all other bins. For third bin we found the the TS value to be 9.19 (2.57 $\sigma$).}
    \label{fig:Figure14}
\end{figure}

The SED, displayed in Figure~\ref{fig:Figure14}, shows a marginal excess in the third energy bin (TS = 9.19), while the remaining bins lie below threshold with TS values of 8.05, 7.83, 7.83, and 7.87. Although the third bin slightly exceeds the TS $>9$ threshold, the absence of consistent excess across other bins and the overall low global TS indicate that this feature is likely a statistical fluctuation rather than genuine emission. A search of the Fermi-LAT 4FGL-DR4 catalog confirms that no previously identified $\gamma$-ray sources exist within $1^\circ$ of VV~114.

\begin{figure}[ht]
\centering
\includegraphics[width=0.6\linewidth]{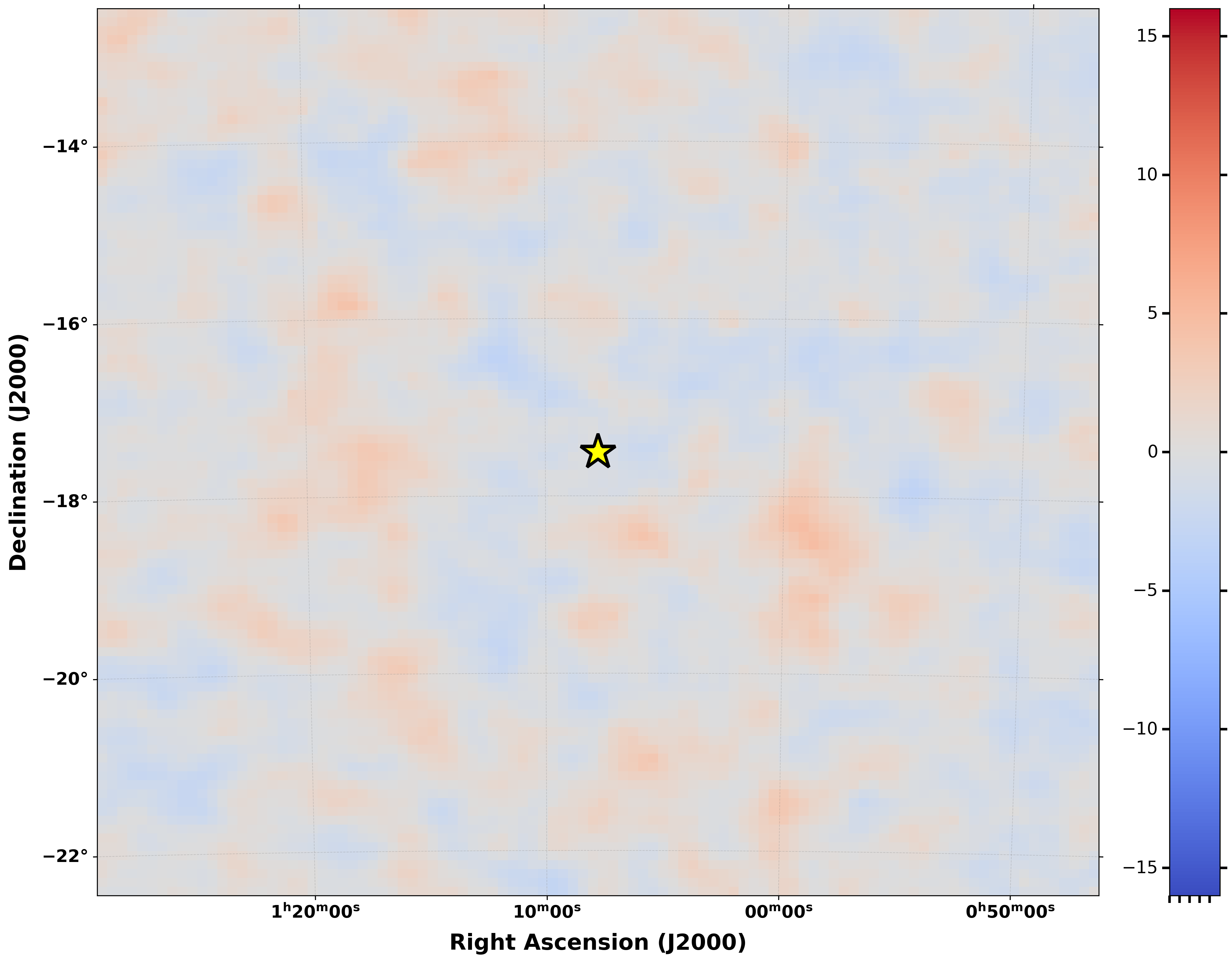}
\caption{\rthis{Residual map of the VV~114 field in the 1--300~GeV range. The plotting conventions are the same as in Figure~\ref{fig:Figure1}.}}
\label{fig:Figure15}
\end{figure}

\rthis{The residual TS map (Figure~\ref{fig:Figure15}) was examined to assess the quality of the spatial fit and to test for potential overfitting or underfitting of the background. The residuals are distributed symmetrically around zero, with a TS range of $-3.22$ to $+4.67$, a mean value of $-0.02$, and a standard deviation of $1.08$. No pixels exceed $\pm3\sigma$, confirming that the model adequately accounts for both diffuse and point-source contributions within the region of interest. The absence of any localized positive or negative residuals indicates that the background model is statistically consistent with the data and that no additional unmodeled $\gamma$-ray component is present near VV~114.}

\rthis{Recent observations with the James Webb Space Telescope (JWST) under the GOALS-JWST program~\citep{Evans2022,Rich2023} provide unprecedented detail on VV~114. Mid-infrared imaging and spectroscopy reveal nearly forty compact knots of intense star formation and a substantial population of previously obscured, massive, dusty star clusters, significantly increasing the known number of active regions. Spatially resolved spectroscopy has clarified the presence of an active galactic nucleus (AGN) in the southwestern nucleus of VV~114E, while confirming vigorous star formation in the northeastern core. These observations offer new insights into the interplay between star formation and possible AGN activity during early-stage galaxy merging~\citep{Evans2022,Rich2023,Gonzalez2024}.}

\section{Control Field Analysis}
\label{sec:control}

\begin{figure*}
    \centering
    \includegraphics[width=0.65\textwidth]{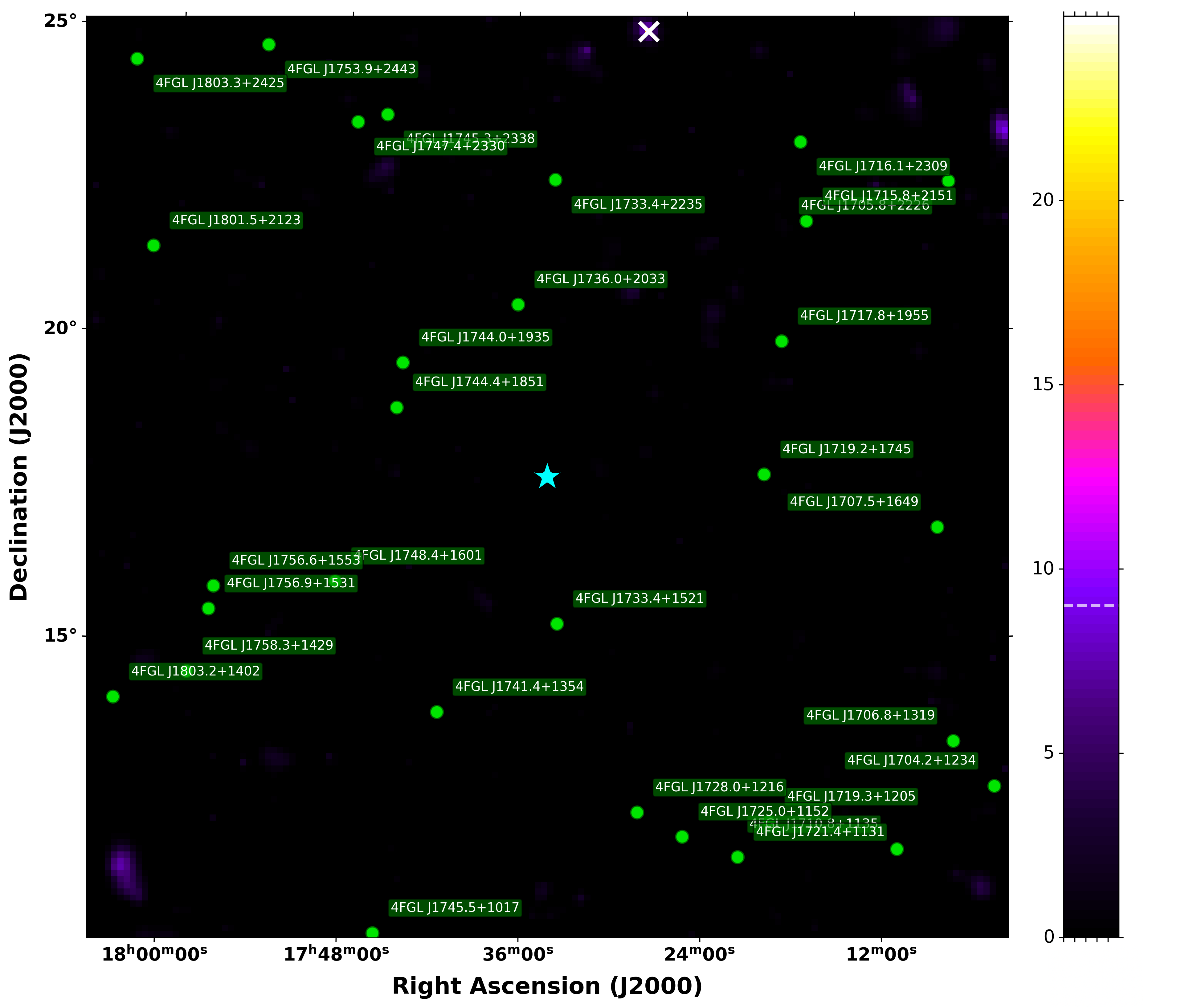}
    \caption{\rthis{TS map of the control field at ($\alpha$, $\delta$) = (263.51$^\circ$, 17.75$^\circ$), located $>2^\circ$ from any 4FGL-DR4 source and at $|b| > 10^\circ$. The TS map, generated using the same energy range (1--300~GeV), time interval, and analysis procedure, shows no significant excess at the center, consistent with background fluctuations. See Section~\ref{sec:control} for details. Cyan star marks the optical center and white cross marks the position with maximum TS.}}
    \label{fig:Figure16}
\end{figure*}

\rthis{To assess the robustness of our $\gamma$-ray detection methodology and to exclude the possibility of spurious or systematic effects, we performed a control field analysis using an empty region of the sky. The control field was centered at ($\alpha$, $\delta$) = (263.51$^\circ$, 17.75$^\circ$), chosen to lie at least $2^\circ$ away from any cataloged 4FGL-DR4 source and to satisfy $|b| > 10^\circ$ in order to minimize contamination from Galactic diffuse emission. No known $\gamma$-ray sources are reported at the control field center. The coordinate was randomly selected to ensure no bias and later checked for any known source in the vicinity.
The control field was analyzed using the exact same configuration as adopted for the science targets, including the energy range (1--300~GeV), ROI size ($10^\circ$ radius), time interval, diffuse background models, and source detection thresholds. This uniform treatment ensures that any differences in the TS distributions between the control and target fields arise solely from genuine astrophysical emission rather than analysis artifacts.
The resulting TS map (Figure~\ref{fig:Figure16}) shows values ranging from TS = 0.00 to 9.96 across the ROI, fully consistent with background statistical fluctuations. The map center exhibits a null TS value, and the maximum TS (TS$_{\mathrm{max}} = 9.96$) occurs at ($\alpha$, $\delta$) = (261.69$^\circ$, 24.99$^\circ$), located $7.44^\circ$ from the ROI center. Only a single pixel reaches TS~$\geq$~9 (corresponding to $\sim$3$\sigma$), and none exceed the canonical detection threshold of TS = 25 ($\sim$5$\sigma$).
This null result confirms that our analysis pipeline does not introduce artificial excesses in regions devoid of real $\gamma$-ray emission. Consequently, the significant $\gamma$-ray signal observed in the merger systems—such as NGC~3256 cannot be attributed to statistical noise or methodological biases. The control field thus provides an empirical baseline demonstrating that our detection procedure is both statistically and systematically reliable.}

\section{Predicted Gamma-ray Fluxes and Comparison with Observations}
\label{sec:flux_comparison}

\rthis{To interpret the observed gamma-ray emission from the analyzed galaxy merger systems, we estimate the expected gamma-ray output based on the empirical correlation between the star formation rate (SFR) and gamma-ray luminosity established for nearby star-forming galaxies by \citet{Ackermann2012}. The relation connects the integrated 0.1--100~GeV luminosity to the SFR as~\citep{Ackermann2012}:}
\begin{equation}
L_{0.1-100\,\mathrm{GeV}} = 1.3\times10^{39}
\left(\frac{\mathrm{SFR}}{M_\odot\,\mathrm{yr^{-1}}}\right)^{1.16}\ \mathrm{erg\,s^{-1}}.
\label{eq:ackermann_scaling}
\end{equation}
\rthis{Although our analysis probes the 1--300~GeV energy range, slightly above the 0.1--100~GeV band used in Equation~\ref{eq:ackermann_scaling}, we adopt this relation as a reasonable approximation.}

The corresponding gamma-ray fluxes are obtained from
\begin{equation}
F_\gamma = \frac{L_\gamma}{4\pi D^2},
\label{eq:flux_relation}
\end{equation}
\rthis{where $D$ is the luminosity distance in centimeters ($1~\mathrm{Mpc}=3.086\times10^{24}~\mathrm{cm}$). Using the SFR and distance values summarized in Table~\ref{tab:merger_properties}, we compute the order-of-magnitude estimates of $F_\gamma$ for each merger system.}

\begin{table}[htbp]
\centering
\caption{\rthis{Comparison between the predicted gamma-ray fluxes derived from the observed SFR--$L_\gamma$ scaling relation and the measured (or 95\% upper limit) fluxes in the 1--300~GeV energy range.}}
\label{tab:pred_flux}
\begin{tabular}{lccc}
\hline
\textbf{Galaxy} & \textbf{SFR} ($M_\odot\,\mathrm{yr^{-1}}$) & \textbf{Predicted Flux} (erg\,cm$^{-2}$\,s$^{-1}$) & \textbf{Measured Flux} (erg\,cm$^{-2}$\,s$^{-1}$) \\
\hline
NGC~3256 & 50 & $\mathcal{O}(10^{-11})$ & $3.8\times10^{-10}$ \\
NGC~660  & 14 & $\mathcal{O}(10^{-10})$ & $1.8\times10^{-10}$ \\
UGC~12914/12915 & 22 & $\mathcal{O}(10^{-11})$ & $<1.9\times10^{-10}$ \\
UGC~813/816  & 2  & $\mathcal{O}(10^{-13})$ & $<3.6\times10^{-10}$ \\
VV~114 & 2.5 & $\mathcal{O}(10^{-13})$ & $<1.3\times10^{-10}$ \\
\hline
\end{tabular}
\end{table}

\rthis{The comparison between the predicted and observed fluxes reveals that the mergers NGC~3256 and NGC~660 exhibit gamma-ray fluxes comparable to the predictions from the empirical SFR--$L_\gamma$ scaling relation, within an order of magnitude. This agreement suggests that these systems may be operating close to the so-called \emph{calorimetric limit}, where most of the cosmic-ray (CR) energy injected by supernovae is lost to inelastic proton--proton ($pp$) interactions within the dense interstellar medium, producing neutral pions that decay into gamma rays. Their high SFRs ($\sim$50 and $\sim$14~$M_\odot~\mathrm{yr^{-1}}$) and correspondingly large molecular gas masses ($\gtrsim10^{10}~M_\odot$) are consistent with efficient CR confinement and energy loss.}

\rthis{In contrast, the remaining merger systems (UGC~12914/12915, UGC~813/816, and VV~114) display upper limits well above the predicted flux levels. The non-detections are therefore consistent with expectations, as the predicted emission lies below the current sensitivity threshold of the Fermi-LAT for typical exposure times. The significant offset between the predicted and measured values, particularly for the Taffy systems, likely reflects the non-calorimetric nature of these galaxies. In such systems, the cosmic-ray residence time is shorter than the proton--proton collision time ($\tau_{\mathrm{res}} < \tau_{pp}$), leading to efficient CR escape before substantial pion production can occur. Additional suppression may arise from the extended geometry of the collisional bridge regions, where the gas densities are lower than in compact starburst nuclei.}

\rthis{We emphasize that the predicted fluxes rely on several simplifying assumptions:  
(1) the empirical relation in Equation~\ref{eq:ackermann_scaling} is derived for normal and starburst galaxies, and its extrapolation to dynamically disturbed mergers may not be exact;  
(2) variations in the IMF, metallicity, or CR acceleration efficiency could alter the normalization by factors of a few; and  
(3) the difference between the 0.1--100~GeV and 1--300~GeV energy bands introduces an additional $\sim$20--30\% uncertainty depending on the photon index.  
Despite these caveats, the scaling relation provides a useful benchmark for assessing whether the observed gamma-ray emission is consistent with expectations from star-formation-powered processes.
Overall, our analysis indicates that the detected gamma-ray fluxes from NGC~3256 and NGC~660 are broadly consistent with expectations based on their intense star-forming activity, supporting the interpretation that their emission may arise predominantly from hadronic interactions of CRs with interstellar gas. For the remaining mergers, the non-detections are compatible with their comparatively low SFRs and expected non-calorimetric behavior. Deeper LAT observations or next-generation MeV--GeV missions will be required to probe the predicted flux regime for these less active systems.  We also note that the neutrino flux corresponding to these predicted gamma-ray emissions is expected to lie several orders of magnitude below the current IceCube sensitivity, consistent with expectations for starburst and merger systems of this type. The detailed analysis of which is beyond the scope of this paper and shall be done in future analysis.}

\section{Conclusions}
\label{sec:conclusions}

We have conducted a targeted search for $\gamma$-ray emission from five galaxy merger systems—NGC~3256, NGC~660, UGC~813/816, UGC~12914/12915, and VV~114—using 16.9 years of Fermi-LAT observations in the 1--300~GeV energy range. None of the systems exhibit a detection exceeding the canonical $5\sigma$ threshold. Marginal $\gamma$-ray signals are observed from NGC~3256 (TS = 15.4, $\sim 3.51\sigma$) and NGC~660 (TS = 8.16, $\sim 2.39\sigma$), with best-fit power-law spectral indices of $\Gamma = 2.05 \pm 0.27$ and $\Gamma = 2.83 \pm 0.54$, respectively. These results are consistent with expectations for starburst-driven $\gamma$-ray emission dominated by cosmic-ray interactions and pion decay, similar to other known starburst galaxies such as NGC~253 and M82~\citep{Abdo2010,Ackermann2012}. The remaining targets—UGC~813/816, UGC~12914/12915, and VV~114—yield TS values below 9, indicating non-detections, although marginal excesses are noted in individual energy bins for UGC~813/816 (second bin, TS = 10.07) and VV~114 (third bin, TS = 9.19). The derived fluxes and upper limits, summarized in Table~\ref{tab:gamma_summary}, provide important constraints on the $\gamma$-ray luminosity and particle acceleration efficiency in these merger systems.

\rthis{The non-detections and low-significance signals are largely limited by the sensitivity of Fermi-LAT, particularly for more distant or intrinsically faint systems. The inclusion of all 4FGL-DR4 sources in the ROI and iterative modeling of residual excesses ensures that the reported fluxes and upper limits are robust against background systematics. While some neighboring cataloged sources may exhibit variability, our preliminary analysis suggests that such effects are minor for the targets considered; dedicated variability studies will be pursued in future work to quantify any potential influence on flux estimates.}

\rthis{Our results indicate that merger-driven star formation and cosmic-ray acceleration can, in principle, produce detectable $\gamma$-ray emission in select systems, as exemplified by NGC~3256. Multiwavelength evidence, including spatially coincident X-ray emission in NGC~3256 and the dense star formation regions revealed by JWST in VV~114, supports the presence of high-energy activity in these mergers, although it does not guarantee a strong $\gamma$-ray signal detectable with current instruments.}

\rthis{Future investigations will aim to disentangle the relative contributions of star formation, cosmic-ray interactions, and potential AGN activity to the observed high-energy emission. Combining infrared and radio tracers of star formation with extended $\gamma$-ray observations will enable more precise modeling of the emission mechanisms. Additionally, next-generation instruments such as the Cherenkov Telescope Array (CTA) and planned $\gamma$-ray observatories will provide improved sensitivity at higher energies ($>100$~GeV), offering the potential to detect fainter or more energetic emission from these and other merging systems.} 

Overall, our study establishes baseline flux measurements and upper limits for prominent galaxy mergers, providing a foundation for future high-energy studies of starburst-driven $\gamma$-ray production in interacting galaxies.

\section{acknowledgements}
SM gratefully acknowledges the Ministry of Education (MoE), Government of India, for their consistent support through the research fellowship, which has been instrumental in facilitating the successful completion of this work. \rthis{We are grateful to the anonymous referees for constructive comments and useful feedback on this manuscript.}

\bibliographystyle{mn2e}
\bibliography{references}

@ARTICLE{Condon1993,
       author = {{Condon}, J.~J. and {Helou}, G. and {Sanders}, D.~B. and {Soifer}, B.~T.},
        title = "{The ``Taffy'' Galaxies UGC 12914/5}",
      journal = {\aj},
         year = 1993,
        month = may,
       volume = {105},
        pages = {1730},
          doi = {10.1086/116549},
       adsurl = {https://ui.adsabs.harvard.edu/abs/1993AJ....105.1730C}
}

@ARTICLE{Gao2003,
       author = {{Gao}, Yu and {Zhu}, Ming and {Seaquist}, E.~R.},
        title = "{Star Formation Across the Taffy Bridge: UGC 12914/15}",
      journal = {\aj},
         year = 2003,
        month = nov,
       volume = {126},
       number = {5},
        pages = {2171-2184},
          doi = {10.1086/378611},
archivePrefix = {arXiv},
       eprint = {astro-ph/0307490},
 primaryClass = {astro-ph},
       adsurl = {https://ui.adsabs.harvard.edu/abs/2003AJ....126.2171G}
}

@ARTICLE{Braine2003,
       author = {{Braine}, J. and {Davoust}, E. and {Zhu}, M. and {Lisenfeld}, U. and {Motch}, C. and {Seaquist}, E.~R.},
        title = "{A molecular gas bridge between the Taffy galaxies}",
      journal = {\aap},
         year = 2003,
        month = sep,
       volume = {408},
        pages = {L13-L16},
          doi = {10.1051/0004-6361:20031148},
archivePrefix = {arXiv},
       eprint = {astro-ph/0308476},
 primaryClass = {astro-ph},
       adsurl = {https://ui.adsabs.harvard.edu/abs/2003A&A...408L..13B}
}

@ARTICLE{Vollmer2012,
       author = {{Vollmer}, B. and {Braine}, J. and {Soida}, M.},
        title = "{A dynamical model for the Taffy galaxies UGC 12914/5}",
      journal = {\aap},
         year = 2012,
        month = nov,
       volume = {547},
          eid = {A39},
        pages = {A39},
          doi = {10.1051/0004-6361/201219668},
archivePrefix = {arXiv},
       eprint = {1209.6052},
 primaryClass = {astro-ph.CO},
       adsurl = {https://ui.adsabs.harvard.edu/abs/2012A&A...547A..39V}
}

@INPROCEEDINGS{Wood2017,
       author = {{Wood}, M. and {Caputo}, R. and {Charles}, E. and {Di Mauro}, M. and {Magill}, J. and {Perkins}, J.~S. and {Fermi-LAT Collaboration}},
        title = "{Fermipy: An open-source Python package for analysis of Fermi-LAT Data}",
    booktitle = {35th International Cosmic Ray Conference (ICRC2017)},
         year = 2017,
       series = {International Cosmic Ray Conference},
       volume = {301},
        month = jul,
          eid = {824},
        pages = {824},
          doi = {10.22323/1.301.0824},
archivePrefix = {arXiv},
       eprint = {1707.09551},
 primaryClass = {astro-ph.IM},
       adsurl = {https://ui.adsabs.harvard.edu/abs/2017ICRC...35..824W}
}

@ARTICLE{Atwood2009,
       author = {{Atwood}, W.~B. and {Abdo}, A.~A. and {Ackermann}, M. and {Althouse}, W. and {Anderson}, B. and {Axelsson}, M. and {Baldini}, L. and {Ballet}, J. and {Band}, D.~L. and {Barbiellini}, G. and {Bartelt}, J. and {Bastieri}, D. and {Baughman}, B.~M. and {Bechtol}, K. and {B{\'e}d{\'e}r{\`e}de}, D. and {Bellardi}, F. and {Bellazzini}, R. and {Berenji}, B. and {Bignami}, G.~F. and {Bisello}, D. and {Bissaldi}, E. and {Blandford}, R.~D. and {Bloom}, E.~D. and {Bogart}, J.~R. and {Bonamente}, E. and {Bonnell}, J. and {Borgland}, A.~W. and {Bouvier}, A. and {Bregeon}, J. and {Brez}, A. and {Brigida}, M. and {Bruel}, P. and {Burnett}, T.~H. and {Busetto}, G. and {Caliandro}, G.~A. and {Cameron}, R.~A. and {Caraveo}, P.~A. and {Carius}, S. and {Carlson}, P. and {Casandjian}, J.~M. and {Cavazzuti}, E. and {Ceccanti}, M. and {Cecchi}, C. and {Charles}, E. and {Chekhtman}, A. and {Cheung}, C.~C. and {Chiang}, J. and {Chipaux}, R. and {Cillis}, A.~N. and {Ciprini}, S. and {Claus}, R. and {Cohen-Tanugi}, J. and {Condamoor}, S. and {Conrad}, J. and {Corbet}, R. and {Corucci}, L. and {Costamante}, L. and {Cutini}, S. and {Davis}, D.~S. and {Decotigny}, D. and {DeKlotz}, M. and {Dermer}, C.~D. and {de Angelis}, A. and {Digel}, S.~W. and {do Couto e Silva}, E. and {Drell}, P.~S. and {Dubois}, R. and {Dumora}, D. and {Edmonds}, Y. and {Fabiani}, D. and {Farnier}, C. and {Favuzzi}, C. and {Flath}, D.~L. and {Fleury}, P. and {Focke}, W.~B. and {Funk}, S. and {Fusco}, P. and {Gargano}, F. and {Gasparrini}, D. and {Gehrels}, N. and {Gentit}, F. -X. and {Germani}, S. and {Giebels}, B. and {Giglietto}, N. and {Giommi}, P. and {Giordano}, F. and {Glanzman}, T. and {Godfrey}, G. and {Grenier}, I.~A. and {Grondin}, M. -H. and {Grove}, J.~E. and {Guillemot}, L. and {Guiriec}, S. and {Haller}, G. and {Harding}, A.~K. and {Hart}, P.~A. and {Hays}, E. and {Healey}, S.~E. and {Hirayama}, M. and {Hjalmarsdotter}, L. and {Horn}, R. and {Hughes}, R.~E. and {J{\'o}hannesson}, G. and {Johansson}, G. and {Johnson}, A.~S. and {Johnson}, R.~P. and {Johnson}, T.~J. and {Johnson}, W.~N. and {Kamae}, T. and {Katagiri}, H. and {Kataoka}, J. and {Kavelaars}, A. and {Kawai}, N. and {Kelly}, H. and {Kerr}, M. and {Klamra}, W. and {Kn{\"o}dlseder}, J. and {Kocian}, M.~L. and {Komin}, N. and {Kuehn}, F. and {Kuss}, M. and {Landriu}, D. and {Latronico}, L. and {Lee}, B. and {Lee}, S. -H. and {Lemoine-Goumard}, M. and {Lionetto}, A.~M. and {Longo}, F. and {Loparco}, F. and {Lott}, B. and {Lovellette}, M.~N. and {Lubrano}, P. and {Madejski}, G.~M. and {Makeev}, A. and {Marangelli}, B. and {Massai}, M.~M. and {Mazziotta}, M.~N. and {McEnery}, J.~E. and {Menon}, N. and {Meurer}, C. and {Michelson}, P.~F. and {Minuti}, M. and {Mirizzi}, N. and {Mitthumsiri}, W. and {Mizuno}, T. and {Moiseev}, A.~A. and {Monte}, C. and {Monzani}, M.~E. and {Moretti}, E. and {Morselli}, A. and {Moskalenko}, I.~V. and {Murgia}, S. and {Nakamori}, T. and {Nishino}, S. and {Nolan}, P.~L. and {Norris}, J.~P. and {Nuss}, E. and {Ohno}, M. and {Ohsugi}, T. and {Omodei}, N. and {Orlando}, E. and {Ormes}, J.~F. and {Paccagnella}, A. and {Paneque}, D. and {Panetta}, J.~H. and {Parent}, D. and {Pearce}, M. and {Pepe}, M. and {Perazzo}, A. and {Pesce-Rollins}, M. and {Picozza}, P. and {Pieri}, L. and {Pinchera}, M. and {Piron}, F. and {Porter}, T.~A. and {Poupard}, L. and {Rain{\`o}}, S. and {Rando}, R. and {Rapposelli}, E. and {Razzano}, M. and {Reimer}, A. and {Reimer}, O. and {Reposeur}, T. and {Reyes}, L.~C. and {Ritz}, S. and {Rochester}, L.~S. and {Rodriguez}, A.~Y. and {Romani}, R.~W. and {Roth}, M. and {Russell}, J.~J. and {Ryde}, F. and {Sabatini}, S. and {Sadrozinski}, H.~F. -W. and {Sanchez}, D. and {Sander}, A. and {Sapozhnikov}, L. and {Parkinson}, P.~M. Saz and {Scargle}, J.~D. and {Schalk}, T.~L. and {Scolieri}, G.},
        title = "{The Large Area Telescope on the Fermi Gamma-Ray Space Telescope Mission}",
      journal = {\apj},
         year = 2009,
        month = jun,
       volume = {697},
       number = {2},
        pages = {1071-1102},
          doi = {10.1088/0004-637X/697/2/1071},
archivePrefix = {arXiv},
       eprint = {0902.1089},
 primaryClass = {astro-ph.IM},
       adsurl = {https://ui.adsabs.harvard.edu/abs/2009ApJ...697.1071A}
}

@ARTICLE{Ballet2023,
       author = {{Ballet}, J. and {Bruel}, P. and {Burnett}, T.~H. and {Lott}, B. and {The Fermi-LAT collaboration}},
        title = "{Fermi Large Area Telescope Fourth Source Catalog Data Release 4 (4FGL-DR4)}",
      journal = {arXiv e-prints},
         year = 2023,
        month = jul,
          eid = {arXiv:2307.12546},
        pages = {arXiv:2307.12546},
          doi = {10.48550/arXiv.2307.12546},
archivePrefix = {arXiv},
       eprint = {2307.12546},
 primaryClass = {astro-ph.HE},
       adsurl = {https://ui.adsabs.harvard.edu/abs/2023arXiv230712546B}
}

@ARTICLE{Zhu2007,
       author = {{Zhu}, Ming and {Gao}, Yu and {Seaquist}, E.~R. and {Dunne}, Loretta},
        title = "{Gas and Dust in the Taffy Galaxies: UGC 12914/15}",
      journal = {\aj},
         year = 2007,
        month = jul,
       volume = {134},
       number = {1},
        pages = {118-134},
          doi = {10.1086/517996},
archivePrefix = {arXiv},
       eprint = {astro-ph/0703200},
 primaryClass = {astro-ph},
       adsurl = {https://ui.adsabs.harvard.edu/abs/2007AJ....134..118Z}
}

@ARTICLE{4FGLDR3,
       author = {{Abdollahi}, S. and {Acero}, F. and {Baldini}, L. and {Ballet}, J. and {Bastieri}, D. and {Bellazzini}, R. and {Berenji}, B. and {Berretta}, A. and {Bissaldi}, E. and {Blandford}, R.~D. and {Bloom}, E. and {Bonino}, R. and {Brill}, A. and {Britto}, R.~J. and {Bruel}, P. and {Burnett}, T.~H. and {Buson}, S. and {Cameron}, R.~A. and {Caputo}, R. and {Caraveo}, P.~A. and {Castro}, D. and {Chaty}, S. and {Cheung}, C.~C. and {Chiaro}, G. and {Cibrario}, N. and {Ciprini}, S. and {Coronado-Bl{\'a}zquez}, J. and {Crnogorcevic}, M. and {Cutini}, S. and {D'Ammando}, F. and {De Gaetano}, S. and {Digel}, S.~W. and {Di Lalla}, N. and {Dirirsa}, F. and {Di Venere}, L. and {Dom{\'\i}nguez}, A. and {Fallah Ramazani}, V. and {Fegan}, S.~J. and {Ferrara}, E.~C. and {Fiori}, A. and {Fleischhack}, H. and {Franckowiak}, A. and {Fukazawa}, Y. and {Funk}, S. and {Fusco}, P. and {Galanti}, G. and {Gammaldi}, V. and {Gargano}, F. and {Garrappa}, S. and {Gasparrini}, D. and {Giacchino}, F. and {Giglietto}, N. and {Giordano}, F. and {Giroletti}, M. and {Glanzman}, T. and {Green}, D. and {Grenier}, I.~A. and {Grondin}, M. -H. and {Guillemot}, L. and {Guiriec}, S. and {Gustafsson}, M. and {Harding}, A.~K. and {Hays}, E. and {Hewitt}, J.~W. and {Horan}, D. and {Hou}, X. and {J{\'o}hannesson}, G. and {Karwin}, C. and {Kayanoki}, T. and {Kerr}, M. and {Kuss}, M. and {Landriu}, D. and {Larsson}, S. and {Latronico}, L. and {Lemoine-Goumard}, M. and {Li}, J. and {Liodakis}, I. and {Longo}, F. and {Loparco}, F. and {Lott}, B. and {Lubrano}, P. and {Maldera}, S. and {Malyshev}, D. and {Manfreda}, A. and {Mart{\'\i}-Devesa}, G. and {Mazziotta}, M.~N. and {Mereu}, I. and {Meyer}, M. and {Michelson}, P.~F. and {Mirabal}, N. and {Mitthumsiri}, W. and {Mizuno}, T. and {Moiseev}, A.~A. and {Monzani}, M.~E. and {Morselli}, A. and {Moskalenko}, I.~V. and {Negro}, M. and {Nuss}, E. and {Omodei}, N. and {Orienti}, M. and {Orlando}, E. and {Paneque}, D. and {Pei}, Z. and {Perkins}, J.~S. and {Persic}, M. and {Pesce-Rollins}, M. and {Petrosian}, V. and {Pillera}, R. and {Poon}, H. and {Porter}, T.~A. and {Principe}, G. and {Rain{\`o}}, S. and {Rando}, R. and {Rani}, B. and {Razzano}, M. and {Razzaque}, S. and {Reimer}, A. and {Reimer}, O. and {Reposeur}, T. and {S{\'a}nchez-Conde}, M. and {Saz Parkinson}, P.~M. and {Scotton}, L. and {Serini}, D. and {Sgr{\`o}}, C. and {Siskind}, E.~J. and {Smith}, D.~A. and {Spandre}, G. and {Spinelli}, P. and {Sueoka}, K. and {Suson}, D.~J. and {Tajima}, H. and {Tak}, D. and {Thayer}, J.~B. and {Thompson}, D.~J. and {Torres}, D.~F. and {Troja}, E. and {Valverde}, J. and {Wood}, K. and {Zaharijas}, G.},
        title = "{Incremental Fermi Large Area Telescope Fourth Source Catalog}",
      journal = {\apjs},
         year = 2022,
        month = jun,
       volume = {260},
       number = {2},
          eid = {53},
        pages = {53},
          doi = {10.3847/1538-4365/ac6751},
archivePrefix = {arXiv},
       eprint = {2201.11184},
 primaryClass = {astro-ph.HE},
       adsurl = {https://ui.adsabs.harvard.edu/abs/2022ApJS..260...53A}
}

@ARTICLE{deMenezes2022,
       author = {{de Menezes}, R.},
        title = "{easyFermi: A graphical interface for performing Fermi-LAT data analyses}",
      journal = {Astronomy and Computing},
         year = 2022,
        month = jul,
       volume = {40},
          eid = {100609},
        pages = {100609},
          doi = {10.1016/j.ascom.2022.100609},
archivePrefix = {arXiv},
       eprint = {2206.11272},
 primaryClass = {astro-ph.HE},
       adsurl = {https://ui.adsabs.harvard.edu/abs/2022A&C....4000609D}
}

@ARTICLE{Donath2023,
       author = {{Donath}, Axel and {Terrier}, R{\'e}gis and {Remy}, Quentin and {Sinha}, Atreyee and {Nigro}, Cosimo and {Pintore}, Fabio and {Kh{\'e}lifi}, Bruno and {Olivera-Nieto}, Laura and {Ruiz}, Jose Enrique and {Br{\"u}gge}, Kai and {Linhoff}, Maximilian and {Contreras}, Jose Luis and {Acero}, Fabio and {Aguasca-Cabot}, Arnau and {Berge}, David and {Bhattacharjee}, Pooja and {Buchner}, Johannes and {Boisson}, Catherine and {Carreto Fidalgo}, David and {Chen}, Andrew and {de Bony de Lavergne}, Mathieu and {de Miranda Cardoso}, Jos{\'e} Vinicius and {Deil}, Christoph and {F{\"u}{\ss}ling}, Matthias and {Funk}, Stefan and {Giunti}, Luca and {Hinton}, Jim and {Jouvin}, L{\'e}a and {King}, Johannes and {Lefaucheur}, Julien and {Lemoine-Goumard}, Marianne and {Lenain}, Jean-Philippe and {L{\'o}pez-Coto}, Rub{\'e}n and {Mohrmann}, Lars and {Morcuende}, Daniel and {Panny}, Sebastian and {Regeard}, Maxime and {Saha}, Lab and {Siejkowski}, Hubert and {Siemiginowska}, Aneta and {Sip{\H{o}}cz}, Brigitta M. and {Unbehaun}, Tim and {van Eldik}, Christopher and {Vuillaume}, Thomas and {Zanin}, Roberta},
        title = "{Gammapy: A Python package for gamma-ray astronomy}",
      journal = {\aap},
         year = 2023,
        month = oct,
       volume = {678},
          eid = {A157},
        pages = {A157},
          doi = {10.1051/0004-6361/202346488},
archivePrefix = {arXiv},
       eprint = {2308.13584},
 primaryClass = {astro-ph.IM},
       adsurl = {https://ui.adsabs.harvard.edu/abs/2023A&A...678A.157D}
}

@ARTICLE{Astropy2018,
       author = {{Astropy Collaboration} and {Price-Whelan}, A.~M. and {Sip{\H{o}}cz}, B.~M. and {G{\"u}nther}, H.~M. and {Lim}, P.~L. and {Crawford}, S.~M. and {Conseil}, S. and {Shupe}, D.~L. and {Craig}, M.~W. and {Dencheva}, N. and {Ginsburg}, A. and {VanderPlas}, J.~T. and {Bradley}, L.~D. and {P{\'e}rez-Su{\'a}rez}, D. and {de Val-Borro}, M. and {Aldcroft}, T.~L. and {Cruz}, K.~L. and {Robitaille}, T.~P. and {Tollerud}, E.~J. and {Ardelean}, C. and {Babej}, T. and {Bach}, Y.~P. and {Bachetti}, M. and {Bakanov}, A.~V. and {Bamford}, S.~P. and {Barentsen}, G. and {Barmby}, P. and {Baumbach}, A. and {Berry}, K.~L. and {Biscani}, F. and {Boquien}, M. and {Bostroem}, K.~A. and {Bouma}, L.~G. and {Brammer}, G.~B. and {Bray}, E.~M. and {Breytenbach}, H. and {Buddelmeijer}, H. and {Burke}, D.~J. and {Calderone}, G. and {Cano Rodr{\'\i}guez}, J.~L. and {Cara}, M. and {Cardoso}, J.~V.~M. and {Cheedella}, S. and {Copin}, Y. and {Corrales}, L. and {Crichton}, D. and {D'Avella}, D. and {Deil}, C. and {Depagne}, {\'E}. and {Dietrich}, J.~P. and {Donath}, A. and {Droettboom}, M. and {Earl}, N. and {Erben}, T. and {Fabbro}, S. and {Ferreira}, L.~A. and {Finethy}, T. and {Fox}, R.~T. and {Garrison}, L.~H. and {Gibbons}, S.~L.~J. and {Goldstein}, D.~A. and {Gommers}, R. and {Greco}, J.~P. and {Greenfield}, P. and {Groener}, A.~M. and {Grollier}, F. and {Hagen}, A. and {Hirst}, P. and {Homeier}, D. and {Horton}, A.~J. and {Hosseinzadeh}, G. and {Hu}, L. and {Hunkeler}, J.~S. and {Ivezi{\'c}}, {\v{Z}}. and {Jain}, A. and {Jenness}, T. and {Kanarek}, G. and {Kendrew}, S. and {Kern}, N.~S. and {Kerzendorf}, W.~E. and {Khvalko}, A. and {King}, J. and {Kirkby}, D. and {Kulkarni}, A.~M. and {Kumar}, A. and {Lee}, A. and {Lenz}, D. and {Littlefair}, S.~P. and {Ma}, Z. and {Macleod}, D.~M. and {Mastropietro}, M. and {McCully}, C. and {Montagnac}, S. and {Morris}, B.~M. and {Mueller}, M. and {Mumford}, S.~J. and {Muna}, D. and {Murphy}, N.~A. and {Nelson}, S. and {Nguyen}, G.~H. and {Ninan}, J.~P. and {N{\"o}the}, M. and {Ogaz}, S. and {Oh}, S. and {Parejko}, J.~K. and {Parley}, N. and {Pascual}, S. and {Patil}, R. and {Patil}, A.~A. and {Plunkett}, A.~L. and {Prochaska}, J.~X. and {Rastogi}, T. and {Reddy Janga}, V. and {Sabater}, J. and {Sakurikar}, P. and {Seifert}, M. and {Sherbert}, L.~E. and {Sherwood-Taylor}, H. and {Shih}, A.~Y. and {Sick}, J. and {Silbiger}, M.~T. and {Singanamalla}, S. and {Singer}, L.~P. and {Sladen}, P.~H. and {Sooley}, K.~A. and {Sornarajah}, S. and {Streicher}, O. and {Teuben}, P. and {Thomas}, S.~W. and {Tremblay}, G.~R. and {Turner}, J.~E.~H. and {Terr{\'o}n}, V. and {van Kerkwijk}, M.~H. and {de la Vega}, A. and {Watkins}, L.~L. and {Weaver}, B.~A. and {Whitmore}, J.~B. and {Woillez}, J. and {Zabalza}, V. and {Astropy Contributors}},
        title = "{The Astropy Project: Building an Open-science Project and Status of the v2.0 Core Package}",
      journal = {\aj},
         year = 2018,
        month = sep,
       volume = {156},
       number = {3},
          eid = {123},
        pages = {123},
          doi = {10.3847/1538-3881/aabc4f},
archivePrefix = {arXiv},
       eprint = {1801.02634},
 primaryClass = {astro-ph.IM},
       adsurl = {https://ui.adsabs.harvard.edu/abs/2018AJ....156..123A}
}

@ARTICLE{Michiyama2018,
       author = {{Michiyama}, Tomonari and {Iono}, Daisuke and {Sliwa}, Kazimierz and {Bolatto}, Alberto and {Nakanishi}, Kouichiro and {Ueda}, Junko and {Saito}, Toshiki and {Ando}, Misaki and {Yamashita}, Takuji and {Yun}, Min},
        title = "{ALMA Observations of HCN and HCO$^{+}$ Outflows in the Merging Galaxy NGC 3256}",
      journal = {\apj},
         year = 2018,
        month = dec,
       volume = {868},
       number = {2},
          eid = {95},
        pages = {95},
          doi = {10.3847/1538-4357/aae82a},
archivePrefix = {arXiv},
       eprint = {1810.04821},
 primaryClass = {astro-ph.GA},
       adsurl = {https://ui.adsabs.harvard.edu/abs/2018ApJ...868...95M}
}

@ARTICLE{Yu2022,
       author = {{Yu}, Niankun and {Ho}, Luis C. and {Wang}, Jing and {Li}, Hangyuan},
        title = "{Statistical Analysis of H I Profile Asymmetry and Shape for Nearby Galaxies}",
      journal = {\apjs},
         year = 2022,
        month = aug,
       volume = {261},
       number = {2},
          eid = {21},
        pages = {21},
          doi = {10.3847/1538-4365/ac626b},
archivePrefix = {arXiv},
       eprint = {2203.13404},
 primaryClass = {astro-ph.GA},
       adsurl = {https://ui.adsabs.harvard.edu/abs/2022ApJS..261...21Y}
}

@ARTICLE{Salter2024,
       author = {{Salter}, C.~J. and {Ghosh}, T. and {Minchin}, R.~F. and {Momjian}, E. and {Catinella}, B. and {Lebron}, M. and {Lerner}, M.~S.},
        title = "{The Discovery and Evolution of a Radio Continuum and Excited-OH Spectral-line Outburst in the Nearby Galaxy NGC 660}",
      journal = {\aj},
         year = 2024,
        month = dec,
       volume = {168},
       number = {6},
          eid = {257},
        pages = {257},
          doi = {10.3847/1538-3881/ad812d},
archivePrefix = {arXiv},
       eprint = {2409.19183},
 primaryClass = {astro-ph.GA},
       adsurl = {https://ui.adsabs.harvard.edu/abs/2024AJ....168..257S}
}

@ARTICLE{Whitmore1990,
       author = {{Whitmore}, Bradley C. and {Lucas}, Ray A. and {McElroy}, Douglas B. and {Steiman-Cameron}, Thomas Y. and {Sackett}, Penny D. and {Olling}, Rob P.},
        title = "{New Observations and a Photographic Atlas of Polar-Ring Galaxies}",
      journal = {\aj},
         year = 1990,
        month = nov,
       volume = {100},
        pages = {1489},
          doi = {10.1086/115614},
       adsurl = {https://ui.adsabs.harvard.edu/abs/1990AJ....100.1489W}
}

@ARTICLE{Springbob2009,
       author = {{Springob}, Christopher M. and {Masters}, Karen L. and {Haynes}, Martha P. and {Giovanelli}, Riccardo and {Marinoni}, Christian},
        title = "{Erratum: ``SFI++ II: A New I-Band Tully-Fisher Catalog, Derivation of Peculiar Velocities and Data Set Properties'' <A href=``bib\_query?2007ApJS..172..599S''> (2007, ApJS, 172, 599)</A>}",
      journal = {\apjs},
         year = 2009,
        month = may,
       volume = {182},
       number = {1},
        pages = {474-475},
          doi = {10.1088/0067-0049/182/1/474},
archivePrefix = {arXiv},
       eprint = {0705.0647},
 primaryClass = {astro-ph},
       adsurl = {https://ui.adsabs.harvard.edu/abs/2009ApJS..182..474S}
}

@ARTICLE{Combes1992,
       author = {{Combes}, F. and {Braine}, J. and {Casoli}, F. and {Gerin}, M. and {van Driel}, W.},
        title = "{Molecular clouds in a polar ring.}",
      journal = {\aap},
         year = 1992,
        month = jun,
       volume = {259},
        pages = {L65-L68},
       adsurl = {https://ui.adsabs.harvard.edu/abs/1992A&A...259L..65C}
}

@ARTICLE{Nazri2021,
       author = {{Nazri}, Nurnabilah and {Annuar}, Adlyka},
        title = "{X-ray source population in the polar ring galaxy NGC 660 as observed by Chandra}",
      journal = {Research in Astronomy and Astrophysics},
         year = 2021,
        month = dec,
       volume = {21},
       number = {11},
          eid = {289},
        pages = {289},
          doi = {10.1088/1674-4527/21/11/289},
       adsurl = {https://ui.adsabs.harvard.edu/abs/2021RAA....21..289N}
}

@ARTICLE{Sakamoto2014,
       author = {{Sakamoto}, Kazushi and {Aalto}, Susanne and {Combes}, Francoise and {Evans}, Aaron and {Peck}, Alison},
        title = "{An Infrared-luminous Merger with Two Bipolar Molecular Outflows: ALMA and SMA Observations of NGC 3256}",
      journal = {\apj},
         year = 2014,
        month = dec,
       volume = {797},
       number = {2},
          eid = {90},
        pages = {90},
          doi = {10.1088/0004-637X/797/2/90},
archivePrefix = {arXiv},
       eprint = {1403.7117},
 primaryClass = {astro-ph.GA},
       adsurl = {https://ui.adsabs.harvard.edu/abs/2014ApJ...797...90S}
}

@ARTICLE{Foreman2013,
       author = {{Foreman-Mackey}, Daniel and {Hogg}, David W. and {Lang}, Dustin and {Goodman}, Jonathan},
        title = "{emcee: The MCMC Hammer}",
      journal = {\pasp},
         year = 2013,
        month = mar,
       volume = {125},
       number = {925},
        pages = {306},
          doi = {10.1086/670067},
archivePrefix = {arXiv},
       eprint = {1202.3665},
 primaryClass = {astro-ph.IM},
       adsurl = {https://ui.adsabs.harvard.edu/abs/2013PASP..125..306F}
}

@ARTICLE{Bushouse1987,
       author = {{Bushouse}, Howard A.},
        title = "{Global Properties of Interacting Disk-Type Galaxies}",
      journal = {\apj},
         year = 1987,
        month = sep,
       volume = {320},
        pages = {49},
          doi = {10.1086/165523},
       adsurl = {https://ui.adsabs.harvard.edu/abs/1987ApJ...320...49B}
}

@ARTICLE{Manna25,
       author = {{Manna}, Siddhant and {Desai}, Shantanu},
        title = "{Search for spatial coincidences between galaxy mergers and Fermi-LAT 4FGL-DR4 sources}",
      journal = {Journal of High Energy Astrophysics},
         year = 2026,
        month = jan,
       volume = {49},
          eid = {100460},
        pages = {100460},
          doi = {10.1016/j.jheap.2025.100460},
archivePrefix = {arXiv},
       eprint = {2507.03970},
 primaryClass = {astro-ph.HE},
       adsurl = {https://ui.adsabs.harvard.edu/abs/2026JHEAp..4900460M}
}

@ARTICLE{Vyaas,
       author = {{Ramakrishnan}, Vyaas and {Desai}, Shantanu},
        title = "{Search for transient gamma-ray emission from magnetar flares using Fermi-LAT}",
      journal = {\jcap},
         year = 2025,
        month = jul,
       volume = {2025},
       number = {7},
          eid = {050},
        pages = {050},
          doi = {10.1088/1475-7516/2025/07/050},
archivePrefix = {arXiv},
       eprint = {2412.03900},
 primaryClass = {astro-ph.HE},
       adsurl = {https://ui.adsabs.harvard.edu/abs/2025JCAP...07..050R}
}

@ARTICLE{Pasumarti,
       author = {{Pasumarti}, Vibhavasu and {Desai}, Shantanu},
        title = "{A study of gamma-ray emission from OJ 287 using Fermi-LAT from 2015-2023}",
      journal = {The Open Journal of Astrophysics},
         year = 2024,
        month = jul,
       volume = {7},
          eid = {64},
        pages = {64},
          doi = {10.33232/001c.121908},
archivePrefix = {arXiv},
       eprint = {2405.15213},
 primaryClass = {astro-ph.HE},
       adsurl = {https://ui.adsabs.harvard.edu/abs/2024OJAp....7E..64P}
}

@ARTICLE{Condon2002,
       author = {{Condon}, J.~J. and {Helou}, G. and {Jarrett}, T.~H.},
        title = "{A Second ``Taffy'' Galaxy Pair}",
      journal = {\aj},
         year = 2002,
        month = apr,
       volume = {123},
       number = {4},
        pages = {1881-1891},
          doi = {10.1086/339558},
       adsurl = {https://ui.adsabs.harvard.edu/abs/2002AJ....123.1881C}
}

@ARTICLE{Braine2004,
       author = {{Braine}, J. and {Lisenfeld}, U. and {Duc}, P. -A. and {Brinks}, E. and {Charmandaris}, V. and {Leon}, S.},
        title = "{Colliding molecular clouds in head-on galaxy collisions}",
      journal = {\aap},
         year = 2004,
        month = may,
       volume = {418},
        pages = {419-428},
          doi = {10.1051/0004-6361:20035732},
archivePrefix = {arXiv},
       eprint = {astro-ph/0402148},
 primaryClass = {astro-ph},
       adsurl = {https://ui.adsabs.harvard.edu/abs/2004A&A...418..419B}
}

@ARTICLE{Knop1994,
       author = {{Knop}, R.~A. and {Soifer}, B.~T. and {Graham}, J.~R. and {Matthews}, K. and {Sanders}, D.~B. and {Scoville}, N.~Z.},
        title = "{VV 114, High Infrared Luminosity Interacting Galaxy System}",
      journal = {\aj},
         year = 1994,
        month = mar,
       volume = {107},
        pages = {920},
          doi = {10.1086/116906},
       adsurl = {https://ui.adsabs.harvard.edu/abs/1994AJ....107..920K}
}

@ARTICLE{Yun1994,
       author = {{Yun}, M.~S. and {Scoville}, N.~Z. and {Knop}, R.~A.},
        title = "{VV 114: Making of an Ultraluminous Galaxy?}",
      journal = {\apjl},
         year = 1994,
        month = aug,
       volume = {430},
        pages = {L109},
          doi = {10.1086/187450},
       adsurl = {https://ui.adsabs.harvard.edu/abs/1994ApJ...430L.109Y}
}

@ARTICLE{Abdo2010,
       author = {{Abdo}, A.~A. and {Ackermann}, M. and {Ajello}, M. and {Atwood}, W.~B. and {Axelsson}, M. and {Baldini}, L. and {Ballet}, J. and {Barbiellini}, G. and {Baring}, M.~G. and {Bastieri}, D. and {Baughman}, B.~M. and {Bechtol}, K. and {Bellazzini}, R. and {Berenji}, B. and {Blandford}, R.~D. and {Bloom}, E.~D. and {Bonamente}, E. and {Borgland}, A.~W. and {Bregeon}, J. and {Brez}, A. and {Brigida}, M. and {Bruel}, P. and {Burnett}, T.~H. and {Buson}, S. and {Caliandro}, G.~A. and {Cameron}, R.~A. and {Camilo}, F. and {Caraveo}, P.~A. and {Casandjian}, J.~M. and {Cecchi}, C. and {{\c{C}}elik}, {\"O}. and {Charles}, E. and {Chekhtman}, A. and {Cheung}, C.~C. and {Chiang}, J. and {Ciprini}, S. and {Claus}, R. and {Cognard}, I. and {Cohen-Tanugi}, J. and {Cominsky}, L.~R. and {Conrad}, J. and {Corbet}, R. and {Cutini}, S. and {den Hartog}, P.~R. and {Dermer}, C.~D. and {de Angelis}, A. and {de Luca}, A. and {de Palma}, F. and {Digel}, S.~W. and {Dormody}, M. and {Silva}, E. do Couto e. and {Drell}, P.~S. and {Dubois}, R. and {Dumora}, D. and {Espinoza}, C. and {Farnier}, C. and {Favuzzi}, C. and {Fegan}, S.~J. and {Ferrara}, E.~C. and {Focke}, W.~B. and {Fortin}, P. and {Frailis}, M. and {Freire}, P.~C.~C. and {Fukazawa}, Y. and {Funk}, S. and {Fusco}, P. and {Gargano}, F. and {Gasparrini}, D. and {Gehrels}, N. and {Germani}, S. and {Giavitto}, G. and {Giebels}, B. and {Giglietto}, N. and {Giommi}, P. and {Giordano}, F. and {Glanzman}, T. and {Godfrey}, G. and {Gotthelf}, E.~V. and {Grenier}, I.~A. and {Grondin}, M. -H. and {Grove}, J.~E. and {Guillemot}, L. and {Guiriec}, S. and {Gwon}, C. and {Hanabata}, Y. and {Harding}, A.~K. and {Hayashida}, M. and {Hays}, E. and {Hughes}, R.~E. and {Jackson}, M.~S. and {J{\'o}hannesson}, G. and {Johnson}, A.~S. and {Johnson}, R.~P. and {Johnson}, T.~J. and {Johnson}, W.~N. and {Johnston}, S. and {Kamae}, T. and {Kanbach}, G. and {Kaspi}, V.~M. and {Katagiri}, H. and {Kataoka}, J. and {Kawai}, N. and {Kerr}, M. and {Kn{\"o}dlseder}, J. and {Kocian}, M.~L. and {Kramer}, M. and {Kuss}, M. and {Lande}, J. and {Latronico}, L. and {Lemoine-Goumard}, M. and {Livingstone}, M. and {Longo}, F. and {Loparco}, F. and {Lott}, B. and {Lovellette}, M.~N. and {Lubrano}, P. and {Lyne}, A.~G. and {Madejski}, G.~M. and {Makeev}, A. and {Manchester}, R.~N. and {Marelli}, M. and {Mazziotta}, M.~N. and {McConville}, W. and {McEnery}, J.~E. and {McGlynn}, S. and {Meurer}, C. and {Michelson}, P.~F. and {Mineo}, T. and {Mitthumsiri}, W. and {Mizuno}, T. and {Moiseev}, A.~A. and {Monte}, C. and {Monzani}, M.~E. and {Morselli}, A. and {Moskalenko}, I.~V. and {Murgia}, S. and {Nakamori}, T. and {Nolan}, P.~L. and {Norris}, J.~P. and {Noutsos}, A. and {Nuss}, E. and {Ohsugi}, T. and {Omodei}, N. and {Orlando}, E. and {Ormes}, J.~F. and {Ozaki}, M. and {Paneque}, D. and {Panetta}, J.~H. and {Parent}, D. and {Pelassa}, V. and {Pepe}, M. and {Pesce-Rollins}, M. and {Piron}, F. and {Porter}, T.~A. and {Rain{\`o}}, S. and {Rando}, R. and {Ransom}, S.~M. and {Ray}, P.~S. and {Razzano}, M. and {Rea}, N. and {Reimer}, A. and {Reimer}, O. and {Reposeur}, T. and {Ritz}, S. and {Rodriguez}, A.~Y. and {Romani}, R.~W. and {Roth}, M. and {Ryde}, F. and {Sadrozinski}, H.~F. -W. and {Sanchez}, D. and {Sander}, A. and {Saz Parkinson}, P.~M. and {Scargle}, J.~D. and {Schalk}, T.~L. and {Sellerholm}, A. and {Sgr{\`o}}, C. and {Siskind}, E.~J. and {Smith}, D.~A. and {Smith}, P.~D. and {Spandre}, G. and {Spinelli}, P. and {Stappers}, B.~W. and {Starck}, J. -L. and {Striani}, E. and {Strickman}, M.~S. and {Strong}, A.~W. and {Suson}, D.~J. and {Tajima}, H. and {Takahashi}, H. and {Takahashi}, T. and {Tanaka}, T. and {Thayer}, J.~B. and {Thayer}, J.~G. and {Theureau}, G. and {Thompson}, D.~J. and {Thorsett}, S.~E. and {Tibaldo}, L. and {Tibolla}, O. and {Torres}, D.~F. and {Tosti}, G.},
        title = "{The First Fermi Large Area Telescope Catalog of Gamma-ray Pulsars}",
      journal = {\apjs},
         year = 2010,
        month = apr,
       volume = {187},
       number = {2},
        pages = {460-494},
          doi = {10.1088/0067-0049/187/2/460},
archivePrefix = {arXiv},
       eprint = {0910.1608},
 primaryClass = {astro-ph.HE},
       adsurl = {https://ui.adsabs.harvard.edu/abs/2010ApJS..187..460A}
}

@ARTICLE{Ackermann2012,
       author = {{Ackermann}, M. and {Ajello}, M. and {Allafort}, A. and {Baldini}, L. and {Ballet}, J. and {Bastieri}, D. and {Bechtol}, K. and {Bellazzini}, R. and {Berenji}, B. and {Bloom}, E.~D. and {Bonamente}, E. and {Borgland}, A.~W. and {Bouvier}, A. and {Bregeon}, J. and {Brigida}, M. and {Bruel}, P. and {Buehler}, R. and {Buson}, S. and {Caliandro}, G.~A. and {Cameron}, R.~A. and {Caraveo}, P.~A. and {Casandjian}, J.~M. and {Cecchi}, C. and {Charles}, E. and {Chekhtman}, A. and {Cheung}, C.~C. and {Chiang}, J. and {Cillis}, A.~N. and {Ciprini}, S. and {Claus}, R. and {Cohen-Tanugi}, J. and {Conrad}, J. and {Cutini}, S. and {de Palma}, F. and {Dermer}, C.~D. and {Digel}, S.~W. and {Silva}, E. do Couto e. and {Drell}, P.~S. and {Drlica-Wagner}, A. and {Favuzzi}, C. and {Fegan}, S.~J. and {Fortin}, P. and {Fukazawa}, Y. and {Funk}, S. and {Fusco}, P. and {Gargano}, F. and {Gasparrini}, D. and {Germani}, S. and {Giglietto}, N. and {Giordano}, F. and {Glanzman}, T. and {Godfrey}, G. and {Grenier}, I.~A. and {Guiriec}, S. and {Gustafsson}, M. and {Hadasch}, D. and {Hayashida}, M. and {Hays}, E. and {Hughes}, R.~E. and {J{\'o}hannesson}, G. and {Johnson}, A.~S. and {Kamae}, T. and {Katagiri}, H. and {Kataoka}, J. and {Kn{\"o}dlseder}, J. and {Kuss}, M. and {Lande}, J. and {Longo}, F. and {Loparco}, F. and {Lott}, B. and {Lovellette}, M.~N. and {Lubrano}, P. and {Madejski}, G.~M. and {Martin}, P. and {Mazziotta}, M.~N. and {McEnery}, J.~E. and {Michelson}, P.~F. and {Mizuno}, T. and {Monte}, C. and {Monzani}, M.~E. and {Morselli}, A. and {Moskalenko}, I.~V. and {Murgia}, S. and {Nishino}, S. and {Norris}, J.~P. and {Nuss}, E. and {Ohno}, M. and {Ohsugi}, T. and {Okumura}, A. and {Omodei}, N. and {Orlando}, E. and {Ozaki}, M. and {Parent}, D. and {Persic}, M. and {Pesce-Rollins}, M. and {Petrosian}, V. and {Pierbattista}, M. and {Piron}, F. and {Pivato}, G. and {Porter}, T.~A. and {Rain{\`o}}, S. and {Rando}, R. and {Razzano}, M. and {Reimer}, A. and {Reimer}, O. and {Ritz}, S. and {Roth}, M. and {Sbarra}, C. and {Sgr{\`o}}, C. and {Siskind}, E.~J. and {Spandre}, G. and {Spinelli}, P. and {Stawarz}, {\L}ukasz and {Strong}, A.~W. and {Takahashi}, H. and {Tanaka}, T. and {Thayer}, J.~B. and {Tibaldo}, L. and {Tinivella}, M. and {Torres}, D.~F. and {Tosti}, G. and {Troja}, E. and {Uchiyama}, Y. and {Vandenbroucke}, J. and {Vianello}, G. and {Vitale}, V. and {Waite}, A.~P. and {Wood}, M. and {Yang}, Z.},
        title = "{GeV Observations of Star-forming Galaxies with the Fermi Large Area Telescope}",
      journal = {\apj},
         year = 2012,
        month = aug,
       volume = {755},
       number = {2},
          eid = {164},
        pages = {164},
          doi = {10.1088/0004-637X/755/2/164},
archivePrefix = {arXiv},
       eprint = {1206.1346},
 primaryClass = {astro-ph.HE},
       adsurl = {https://ui.adsabs.harvard.edu/abs/2012ApJ...755..164A}
}

@ARTICLE{Kashiyama2014,
       author = {{Kashiyama}, Kazumi and {M{\'e}sz{\'a}ros}, Peter},
        title = "{Galaxy Mergers as a Source of Cosmic Rays, Neutrinos, and Gamma Rays}",
      journal = {\apjl},
         year = 2014,
        month = jul,
       volume = {790},
       number = {1},
          eid = {L14},
        pages = {L14},
          doi = {10.1088/2041-8205/790/1/L14},
archivePrefix = {arXiv},
       eprint = {1405.3262},
 primaryClass = {astro-ph.HE},
       adsurl = {https://ui.adsabs.harvard.edu/abs/2014ApJ...790L..14K}
}

@ARTICLE{Sargent1989,
       author = {{Sargent}, A.~I. and {Sanders}, D.~B. and {Phillips}, T.~G.},
        title = "{CO(21) Emission from the Interacting Galaxy Pair NGC 3256}",
      journal = {\apjl},
         year = 1989,
        month = nov,
       volume = {346},
        pages = {L9},
          doi = {10.1086/185566},
       adsurl = {https://ui.adsabs.harvard.edu/abs/1989ApJ...346L...9S}
}

@ARTICLE{Lira2002,
       author = {{Lira}, P. and {Ward}, M. and {Zezas}, A. and {Alonso-Herrero}, A. and {Ueno}, S.},
        title = "{Chandra observations of the luminous infrared galaxy NGC 3256}",
      journal = {\mnras},
         year = 2002,
        month = feb,
       volume = {330},
       number = {2},
        pages = {259-278},
          doi = {10.1046/j.1365-8711.2002.05014.x},
archivePrefix = {arXiv},
       eprint = {astro-ph/0109198},
 primaryClass = {astro-ph},
       adsurl = {https://ui.adsabs.harvard.edu/abs/2002MNRAS.330..259L}
}

@ARTICLE{Zepf1999,
       author = {{Zepf}, Stephen E. and {Ashman}, Keith M. and {English}, Jayanne and {Freeman}, Kenneth C. and {Sharples}, Ray M.},
        title = "{The Formation and Evolution of Candidate Young Globular Clusters in NGC 3256}",
      journal = {\aj},
         year = 1999,
        month = aug,
       volume = {118},
       number = {2},
        pages = {752-764},
          doi = {10.1086/300961},
archivePrefix = {arXiv},
       eprint = {astro-ph/9904247},
 primaryClass = {astro-ph},
       adsurl = {https://ui.adsabs.harvard.edu/abs/1999AJ....118..752Z}
}

@ARTICLE{Aalto1995,
       author = {{Aalto}, S. and {Booth}, R.~S. and {Black}, J.~H. and {Johansson}, L.~E.~B.},
        title = "{Molecular gas in starburst galaxies: line intensities and physical conditions}",
      journal = {\aap},
         year = 1995,
        month = aug,
       volume = {300},
        pages = {369},
       adsurl = {https://ui.adsabs.harvard.edu/abs/1995A&A...300..369A}
}

@ARTICLE{Vandriel1995,
       author = {{van Driel}, W. and {Combes}, F. and {Casoli}, F. and {Gerin}, M. and {Nakai}, N. and {Miyaji}, T. and {Hamabe}, M. and {Sofue}, Y. and {Ichikawa}, T. and {Yoshida}, S. and {Kobayashi}, Y. and {Geng}, F. and {Minezaki}, T. and {Arimoto}, N. and {Kodama}, T. and {Goudfrooij}, P. and {Mulder}, P.~S. and {Wakamatsu}, K. and {Yanagisawa}, K.},
        title = "{Polar Ring Spiral Galaxy NGC 660}",
      journal = {\aj},
         year = 1995,
        month = mar,
       volume = {109},
        pages = {942},
          doi = {10.1086/117333},
       adsurl = {https://ui.adsabs.harvard.edu/abs/1995AJ....109..942V}
}

@ARTICLE{Barnes1992,
       author = {{Barnes}, Joshua E. and {Hernquist}, Lars},
        title = "{Dynamics of interacting galaxies.}",
      journal = {\araa},
         year = 1992,
        month = jan,
       volume = {30},
        pages = {705-742},
          doi = {10.1146/annurev.aa.30.090192.003421},
       adsurl = {https://ui.adsabs.harvard.edu/abs/1992ARA&A..30..705B}
}

@ARTICLE{Hopkins2006,
       author = {{Hopkins}, Philip F. and {Hernquist}, Lars and {Cox}, Thomas J. and {Di Matteo}, Tiziana and {Robertson}, Brant and {Springel}, Volker},
        title = "{A Unified, Merger-driven Model of the Origin of Starbursts, Quasars, the Cosmic X-Ray Background, Supermassive Black Holes, and Galaxy Spheroids}",
      journal = {\apjs},
         year = 2006,
        month = mar,
       volume = {163},
       number = {1},
        pages = {1-49},
          doi = {10.1086/499298},
archivePrefix = {arXiv},
       eprint = {astro-ph/0506398},
 primaryClass = {astro-ph},
       adsurl = {https://ui.adsabs.harvard.edu/abs/2006ApJS..163....1H}
}

@ARTICLE{Lacki2011,
       author = {{Lacki}, Brian C. and {Thompson}, Todd A. and {Quataert}, Eliot and {Loeb}, Abraham and {Waxman}, Eli},
        title = "{On the GeV and TeV Detections of the Starburst Galaxies M82 and NGC 253}",
      journal = {\apj},
         year = 2011,
        month = jun,
       volume = {734},
       number = {2},
          eid = {107},
        pages = {107},
          doi = {10.1088/0004-637X/734/2/107},
archivePrefix = {arXiv},
       eprint = {1003.3257},
 primaryClass = {astro-ph.HE},
       adsurl = {https://ui.adsabs.harvard.edu/abs/2011ApJ...734..107L}
}

@ARTICLE{Sanders1996,
       author = {{Sanders}, D.~B. and {Mirabel}, I.~F.},
        title = "{Luminous Infrared Galaxies}",
      journal = {\araa},
         year = 1996,
        month = jan,
       volume = {34},
        pages = {749},
          doi = {10.1146/annurev.astro.34.1.749},
       adsurl = {https://ui.adsabs.harvard.edu/abs/1996ARA&A..34..749S}
}

@ARTICLE{Bouri2025,
       author = {{Bouri}, Subhadip and {Parashari}, Priyank and {Das}, Mousumi and {Laha}, Ranjan},
        title = "{First search for high-energy neutrino emission from galaxy mergers}",
      journal = {\prd},
         year = 2025,
        month = mar,
       volume = {111},
       number = {6},
          eid = {063059},
        pages = {063059},
          doi = {10.1103/PhysRevD.111.063059},
archivePrefix = {arXiv},
       eprint = {2404.06539},
 primaryClass = {astro-ph.HE},
       adsurl = {https://ui.adsabs.harvard.edu/abs/2025PhRvD.111f3059B}
}

@ARTICLE{Abdo2009,
       author = {{Abdo}, A.~A. and {Ackermann}, M. and {Ajello}, M. and {Atwood}, W.~B. and {Axelsson}, M. and {Baldini}, L. and {Ballet}, J. and {Band}, D.~L. and {Barbiellini}, G. and {Bastieri}, D. and {Battelino}, M. and {Baughman}, B.~M. and {Bechtol}, K. and {Bellazzini}, R. and {Berenji}, B. and {Bignami}, G.~F. and {Blandford}, R.~D. and {Bloom}, E.~D. and {Bonamente}, E. and {Borgland}, A.~W. and {Bouvier}, A. and {Bregeon}, J. and {Brez}, A. and {Brigida}, M. and {Bruel}, P. and {Burnett}, T.~H. and {Caliandro}, G.~A. and {Cameron}, R.~A. and {Caraveo}, P.~A. and {Casandjian}, J.~M. and {Cavazzuti}, E. and {Cecchi}, C. and {Charles}, E. and {Chekhtman}, A. and {Cheung}, C.~C. and {Chiang}, J. and {Ciprini}, S. and {Claus}, R. and {Cohen-Tanugi}, J. and {Cominsky}, L.~R. and {Conrad}, J. and {Corbet}, R. and {Costamante}, L. and {Cutini}, S. and {Davis}, D.~S. and {Dermer}, C.~D. and {de Angelis}, A. and {de Luca}, A. and {de Palma}, F. and {Digel}, S.~W. and {Dormody}, M. and {do Couto e Silva}, E. and {Drell}, P.~S. and {Dubois}, R. and {Dumora}, D. and {Farnier}, C. and {Favuzzi}, C. and {Fegan}, S.~J. and {Ferrara}, E.~C. and {Focke}, W.~B. and {Frailis}, M. and {Fukazawa}, Y. and {Funk}, S. and {Fusco}, P. and {Gargano}, F. and {Gasparrini}, D. and {Gehrels}, N. and {Germani}, S. and {Giebels}, B. and {Giglietto}, N. and {Giommi}, P. and {Giordano}, F. and {Glanzman}, T. and {Godfrey}, G. and {Grenier}, I.~A. and {Grondin}, M. -H. and {Grove}, J.~E. and {Guillemot}, L. and {Guiriec}, S. and {Hanabata}, Y. and {Harding}, A.~K. and {Hartman}, R.~C. and {Hayashida}, M. and {Hays}, E. and {Healey}, S.~E. and {Horan}, D. and {Hughes}, R.~E. and {J{\'o}hannesson}, G. and {Johnson}, A.~S. and {Johnson}, R.~P. and {Johnson}, T.~J. and {Johnson}, W.~N. and {Kamae}, T. and {Katagiri}, H. and {Kataoka}, J. and {Kawai}, N. and {Kerr}, M. and {Kn{\"o}dlseder}, J. and {Kocevski}, D. and {Kocian}, M.~L. and {Komin}, N. and {Kuehn}, F. and {Kuss}, M. and {Lande}, J. and {Latronico}, L. and {Lee}, S. -H. and {Lemoine-Goumard}, M. and {Longo}, F. and {Loparco}, F. and {Lott}, B. and {Lovellette}, M.~N. and {Lubrano}, P. and {Madejski}, G.~M. and {Makeev}, A. and {Marelli}, M. and {Mazziotta}, M.~N. and {McConville}, W. and {McEnery}, J.~E. and {McGlynn}, S. and {Meurer}, C. and {Michelson}, P.~F. and {Mitthumsiri}, W. and {Mizuno}, T. and {Moiseev}, A.~A. and {Monte}, C. and {Monzani}, M.~E. and {Moretti}, E. and {Morselli}, A. and {Moskalenko}, I.~V. and {Murgia}, S. and {Nakamori}, T. and {Nolan}, P.~L. and {Norris}, J.~P. and {Nuss}, E. and {Ohno}, M. and {Ohsugi}, T. and {Omodei}, N. and {Orlando}, E. and {Ormes}, J.~F. and {Ozaki}, M. and {Paneque}, D. and {Panetta}, J.~H. and {Parent}, D. and {Pelassa}, V. and {Pepe}, M. and {Pesce-Rollins}, M. and {Piron}, F. and {Porter}, T.~A. and {Poupard}, L. and {Rain{\`o}}, S. and {Rando}, R. and {Ray}, P.~S. and {Razzano}, M. and {Rea}, N. and {Reimer}, A. and {Reimer}, O. and {Reposeur}, T. and {Ritz}, S. and {Rochester}, L.~S. and {Rodriguez}, A.~Y. and {Romani}, R.~W. and {Roth}, M. and {Ryde}, F. and {Sadrozinski}, H.~F. -W. and {Sanchez}, D. and {Sander}, A. and {Saz Parkinson}, P.~M. and {Scargle}, J.~D. and {Schalk}, T.~L. and {Sellerholm}, A. and {Sgr{\`o}}, C. and {Shaw}, M.~S. and {Shrader}, C. and {Sierpowska-Bartosik}, A. and {Siskind}, E.~J. and {Smith}, D.~A. and {Smith}, P.~D. and {Spandre}, G. and {Spinelli}, P. and {Starck}, J. -L. and {Stephens}, T.~E. and {Strickman}, M.~S. and {Strong}, A.~W. and {Suson}, D.~J. and {Tajima}, H. and {Takahashi}, H. and {Takahashi}, T. and {Tanaka}, T. and {Thayer}, J.~B. and {Thayer}, J.~G. and {Thompson}, D.~J. and {Tibaldo}, L. and {Tibolla}, O. and {Torres}, D.~F. and {Tosti}, G. and {Tramacere}, A. and {Uchiyama}, Y. and {Usher}, T.~L. and {Van Etten}, A. and {Vilchez}, N.},
        title = "{Fermi/Large Area Telescope Bright Gamma-Ray Source List}",
      journal = {\apjs},
         year = 2009,
        month = jul,
       volume = {183},
       number = {1},
        pages = {46-66},
          doi = {10.1088/0067-0049/183/1/46},
archivePrefix = {arXiv},
       eprint = {0902.1340},
 primaryClass = {astro-ph.HE},
       adsurl = {https://ui.adsabs.harvard.edu/abs/2009ApJS..183...46A}
}

@ARTICLE{Abdollahi2020,
       author = {{Abdollahi}, S. and {Acero}, F. and {Baldini}, L. and {Ballet}, J. and {Bastieri}, D. and {Bellazzini}, R. and {Berenji}, B. and {Berretta}, A. and {Bissaldi}, E. and {Blandford}, R.~D. and {Bloom}, E. and {Bonino}, R. and {Brill}, A. and {Britto}, R.~J. and {Bruel}, P. and {Burnett}, T.~H. and {Buson}, S. and {Cameron}, R.~A. and {Caputo}, R. and {Caraveo}, P.~A. and {Castro}, D. and {Chaty}, S. and {Cheung}, C.~C. and {Chiaro}, G. and {Cibrario}, N. and {Ciprini}, S. and {Coronado-Bl{\'a}zquez}, J. and {Crnogorcevic}, M. and {Cutini}, S. and {D'Ammando}, F. and {De Gaetano}, S. and {Digel}, S.~W. and {Di Lalla}, N. and {Dirirsa}, F. and {Di Venere}, L. and {Dom{\'\i}nguez}, A. and {Fallah Ramazani}, V. and {Fegan}, S.~J. and {Ferrara}, E.~C. and {Fiori}, A. and {Fleischhack}, H. and {Franckowiak}, A. and {Fukazawa}, Y. and {Funk}, S. and {Fusco}, P. and {Galanti}, G. and {Gammaldi}, V. and {Gargano}, F. and {Garrappa}, S. and {Gasparrini}, D. and {Giacchino}, F. and {Giglietto}, N. and {Giordano}, F. and {Giroletti}, M. and {Glanzman}, T. and {Green}, D. and {Grenier}, I.~A. and {Grondin}, M. -H. and {Guillemot}, L. and {Guiriec}, S. and {Gustafsson}, M. and {Harding}, A.~K. and {Hays}, E. and {Hewitt}, J.~W. and {Horan}, D. and {Hou}, X. and {J{\'o}hannesson}, G. and {Karwin}, C. and {Kayanoki}, T. and {Kerr}, M. and {Kuss}, M. and {Landriu}, D. and {Larsson}, S. and {Latronico}, L. and {Lemoine-Goumard}, M. and {Li}, J. and {Liodakis}, I. and {Longo}, F. and {Loparco}, F. and {Lott}, B. and {Lubrano}, P. and {Maldera}, S. and {Malyshev}, D. and {Manfreda}, A. and {Mart{\'\i}-Devesa}, G. and {Mazziotta}, M.~N. and {Mereu}, I. and {Meyer}, M. and {Michelson}, P.~F. and {Mirabal}, N. and {Mitthumsiri}, W. and {Mizuno}, T. and {Moiseev}, A.~A. and {Monzani}, M.~E. and {Morselli}, A. and {Moskalenko}, I.~V. and {Negro}, M. and {Nuss}, E. and {Omodei}, N. and {Orienti}, M. and {Orlando}, E. and {Paneque}, D. and {Pei}, Z. and {Perkins}, J.~S. and {Persic}, M. and {Pesce-Rollins}, M. and {Petrosian}, V. and {Pillera}, R. and {Poon}, H. and {Porter}, T.~A. and {Principe}, G. and {Rain{\`o}}, S. and {Rando}, R. and {Rani}, B. and {Razzano}, M. and {Razzaque}, S. and {Reimer}, A. and {Reimer}, O. and {Reposeur}, T. and {S{\'a}nchez-Conde}, M. and {Saz Parkinson}, P.~M. and {Scotton}, L. and {Serini}, D. and {Sgr{\`o}}, C. and {Siskind}, E.~J. and {Smith}, D.~A. and {Spandre}, G. and {Spinelli}, P. and {Sueoka}, K. and {Suson}, D.~J. and {Tajima}, H. and {Tak}, D. and {Thayer}, J.~B. and {Thompson}, D.~J. and {Torres}, D.~F. and {Troja}, E. and {Valverde}, J. and {Wood}, K. and {Zaharijas}, G.},
        title = "{Incremental Fermi Large Area Telescope Fourth Source Catalog}",
      journal = {\apjs},
         year = 2022,
        month = jun,
       volume = {260},
       number = {2},
          eid = {53},
        pages = {53},
          doi = {10.3847/1538-4365/ac6751},
archivePrefix = {arXiv},
       eprint = {2201.11184},
 primaryClass = {astro-ph.HE},
       adsurl = {https://ui.adsabs.harvard.edu/abs/2022ApJS..260...53A}
}

@ARTICLE{Mattox1996,
       author = {{Mattox}, J.~R. and {Bertsch}, D.~L. and {Chiang}, J. and {Dingus}, B.~L. and {Digel}, S.~W. and {Esposito}, J.~A. and {Fierro}, J.~M. and {Hartman}, R.~C. and {Hunter}, S.~D. and {Kanbach}, G. and {Kniffen}, D.~A. and {Lin}, Y.~C. and {Macomb}, D.~J. and {Mayer-Hasselwander}, H.~A. and {Michelson}, P.~F. and {von Montigny}, C. and {Mukherjee}, R. and {Nolan}, P.~L. and {Ramanamurthy}, P.~V. and {Schneid}, E. and {Sreekumar}, P. and {Thompson}, D.~J. and {Willis}, T.~D.},
        title = "{The Likelihood Analysis of EGRET Data}",
      journal = {\apj},
         year = 1996,
        month = apr,
       volume = {461},
        pages = {396},
          doi = {10.1086/177068},
       adsurl = {https://ui.adsabs.harvard.edu/abs/1996ApJ...461..396M}
}

@ARTICLE{James1975,
       author = {{James}, F. and {Roos}, M.},
        title = "{Minuit - a system for function minimization and analysis of the parameter errors and correlations}",
      journal = {Computer Physics Communications},
         year = 1975,
        month = dec,
       volume = {10},
       number = {6},
        pages = {343-367},
          doi = {10.1016/0010-4655(75)90039-9},
       adsurl = {https://ui.adsabs.harvard.edu/abs/1975CoPhC..10..343J}
}

@ARTICLE{Lira2008,
       author = {{Lira}, P. and {Gonzalez-Corvalan}, V. and {Ward}, M. and {Hoyer}, S.},
        title = "{An infrared study of the double nucleus in NGC3256}",
      journal = {\mnras},
         year = 2008,
        month = feb,
       volume = {384},
       number = {1},
        pages = {316-322},
          doi = {10.1111/j.1365-2966.2007.12705.x},
archivePrefix = {arXiv},
       eprint = {0711.2045},
 primaryClass = {astro-ph},
       adsurl = {https://ui.adsabs.harvard.edu/abs/2008MNRAS.384..316L}
}

@ARTICLE{Atwood2013,
       author = {{Atwood}, W. and {Albert}, A. and {Baldini}, L. and {Tinivella}, M. and {Bregeon}, J. and {Pesce-Rollins}, M. and {Sgr{\`o}}, C. and {Bruel}, P. and {Charles}, E. and {Drlica-Wagner}, A. and {Franckowiak}, A. and {Jogler}, T. and {Rochester}, L. and {Usher}, T. and {Wood}, M. and {Cohen-Tanugi}, J. and {Zimmer}, S.},
        title = "{Pass 8: Toward the Full Realization of the Fermi-LAT Scientific Potential}",
      journal = {arXiv e-prints},
         year = 2013,
        month = mar,
          eid = {arXiv:1303.3514},
        pages = {arXiv:1303.3514},
          doi = {10.48550/arXiv.1303.3514},
archivePrefix = {arXiv},
       eprint = {1303.3514},
 primaryClass = {astro-ph.IM},
       adsurl = {https://ui.adsabs.harvard.edu/abs/2013arXiv1303.3514A}
}

@ARTICLE{Lisenfeld2008,
       author = {{Lisenfeld}, U. and {Mundell}, C.~G. and {Schinnerer}, E. and {Appleton}, P.~N. and {Allsopp}, J.},
        title = "{Molecular Gas and Dust in Arp 94: The Formation of a Recycled Galaxy in an Interacting System}",
      journal = {\apj},
         year = 2008,
        month = sep,
       volume = {685},
       number = {1},
        pages = {181-193},
          doi = {10.1086/590420},
archivePrefix = {arXiv},
       eprint = {0807.0176},
 primaryClass = {astro-ph},
       adsurl = {https://ui.adsabs.harvard.edu/abs/2008ApJ...685..181L}
}

@ARTICLE{Lisenfeld2010,
       author = {{Lisenfeld}, U. and {V{\"o}lk}, H.~J.},
        title = "{Shock acceleration of relativistic particles in galaxy-galaxy collisions}",
      journal = {\aap},
         year = 2010,
        month = dec,
       volume = {524},
          eid = {A27},
        pages = {A27},
          doi = {10.1051/0004-6361/201015083},
archivePrefix = {arXiv},
       eprint = {1009.1659},
 primaryClass = {astro-ph.CO},
       adsurl = {https://ui.adsabs.harvard.edu/abs/2010A&A...524A..27L}
}

@article{Wilks1938,
  title={The large-sample distribution of the likelihood ratio for testing composite hypotheses},
  author={Wilks, Samuel S},
  journal={The annals of mathematical statistics},
  volume={9},
  number={1},
  pages={60--62},
  year={1938},
  publisher={JSTOR}
}

@ARTICLE{Manna2024,
       author = {{Manna}, Siddhant and {Desai}, Shantanu},
        title = "{Search for GeV gamma-ray emission from SPT-SZ selected galaxy clusters with 15 years of Fermi-LAT data}",
      journal = {\jcap},
         year = 2024,
        month = jan,
       volume = {2024},
       number = {1},
          eid = {017},
        pages = {017},
          doi = {10.1088/1475-7516/2024/01/017},
archivePrefix = {arXiv},
       eprint = {2310.07519},
 primaryClass = {astro-ph.HE},
       adsurl = {https://ui.adsabs.harvard.edu/abs/2024JCAP...01..017M}
}

@ARTICLE{Watson2009,
       author = {{Watson}, M.~G. and {Schr{\"o}der}, A.~C. and {Fyfe}, D. and {Page}, C.~G. and {Lamer}, G. and {Mateos}, S. and {Pye}, J. and {Sakano}, M. and {Rosen}, S. and {Ballet}, J. and {Barcons}, X. and {Barret}, D. and {Boller}, T. and {Brunner}, H. and {Brusa}, M. and {Caccianiga}, A. and {Carrera}, F.~J. and {Ceballos}, M. and {Della Ceca}, R. and {Denby}, M. and {Denkinson}, G. and {Dupuy}, S. and {Farrell}, S. and {Fraschetti}, F. and {Freyberg}, M.~J. and {Guillout}, P. and {Hambaryan}, V. and {Maccacaro}, T. and {Mathiesen}, B. and {McMahon}, R. and {Michel}, L. and {Motch}, C. and {Osborne}, J.~P. and {Page}, M. and {Pakull}, M.~W. and {Pietsch}, W. and {Saxton}, R. and {Schwope}, A. and {Severgnini}, P. and {Simpson}, M. and {Sironi}, G. and {Stewart}, G. and {Stewart}, I.~M. and {Stobbart}, A. -M. and {Tedds}, J. and {Warwick}, R. and {Webb}, N. and {West}, R. and {Worrall}, D. and {Yuan}, W.},
        title = "{The XMM-Newton serendipitous survey. V. The Second XMM-Newton serendipitous source catalogue}",
      journal = {\aap},
         year = 2009,
        month = jan,
       volume = {493},
       number = {1},
        pages = {339-373},
          doi = {10.1051/0004-6361:200810534},
archivePrefix = {arXiv},
       eprint = {0807.1067},
 primaryClass = {astro-ph},
       adsurl = {https://ui.adsabs.harvard.edu/abs/2009A&A...493..339W}
}

@ARTICLE{Evans2022,
       author = {{Evans}, A.~S. and {Frayer}, D.~T. and {Charmandaris}, Vassilis and {Armus}, Lee and {Inami}, Hanae and {Surace}, Jason and {Linden}, Sean and {Soifer}, B.~T. and {Diaz-Santos}, Tanio and {Larson}, Kirsten L. and {Rich}, Jeffrey A. and {Song}, Yiqing and {Barcos-Munoz}, Loreto and {Mazzarella}, Joseph M. and {Privon}, George C. and {U}, Vivian and {Medling}, Anne M. and {B{\"o}ker}, Torsten and {Aalto}, Susanne and {Iwasawa}, Kazushi and {Howell}, Justin H. and {van der Werf}, Paul and {Appleton}, Philip and {Bohn}, Thomas and {Brown}, Michael J.~I. and {Hayward}, Christopher C. and {Hoshioka}, Shunshi and {Kemper}, Francisca and {Lai}, Thomas and {Law}, David and {Malkan}, Matthew A. and {Marshall}, Jason and {Murphy}, Eric J. and {Sanders}, David and {Stierwalt}, Sabrina},
        title = "{GOALS-JWST: Hidden Star Formation and Extended PAH Emission in the Luminous Infrared Galaxy VV 114}",
      journal = {\apjl},
         year = 2022,
        month = nov,
       volume = {940},
       number = {1},
          eid = {L8},
        pages = {L8},
          doi = {10.3847/2041-8213/ac9971},
archivePrefix = {arXiv},
       eprint = {2208.14507},
 primaryClass = {astro-ph.GA},
       adsurl = {https://ui.adsabs.harvard.edu/abs/2022ApJ...940L...8E}
}

@ARTICLE{Rich2023,
       author = {{Rich}, J. and {Aalto}, S. and {Evans}, A.~S. and {Charmandaris}, V. and {Privon}, G.~C. and {Lai}, T. and {Inami}, H. and {Linden}, S. and {Armus}, L. and {Diaz-Santos}, T. and {Appleton}, P. and {Barcos-Mu{\~n}oz}, L. and {B{\"o}ker}, T. and {Larson}, K.~L. and {Law}, D.~R. and {Malkan}, M.~A. and {Medling}, A.~M. and {Song}, Y. and {U}, V. and {van der Werf}, P. and {Bohn}, T. and {Brown}, M.~J.~I. and {Finnerty}, L. and {Hayward}, C. and {Howell}, J. and {Iwasawa}, K. and {Kemper}, F. and {Marshall}, J. and {Mazzarella}, J.~M. and {McKinney}, J. and {Muller-Sanchez}, F. and {Murphy}, E.~J. and {Sanders}, D. and {Soifer}, B.~T. and {Stierwalt}, S. and {Surace}, J.},
        title = "{GOALS-JWST: Pulling Back the Curtain on the AGN and Star Formation in VV 114}",
      journal = {\apjl},
         year = 2023,
        month = feb,
       volume = {944},
       number = {2},
          eid = {L50},
        pages = {L50},
          doi = {10.3847/2041-8213/acb2b8},
archivePrefix = {arXiv},
       eprint = {2301.02338},
 primaryClass = {astro-ph.GA},
       adsurl = {https://ui.adsabs.harvard.edu/abs/2023ApJ...944L..50R}
}

@ARTICLE{Gonzalez2024,
       author = {{Gonz{\'a}lez-Alfonso}, Eduardo and {Garc{\'\i}a-Bernete}, Ismael and {Pereira-Santaella}, Miguel and {Neufeld}, David A. and {Fischer}, Jacqueline and {Donnan}, Fergus R.},
        title = "{JWST detection of extremely excited outflowing CO and H$_{2}$O in VV 114 E SW: A possible rapidly accreting IMBH}",
      journal = {\aap},
         year = 2024,
        month = feb,
       volume = {682},
          eid = {A182},
        pages = {A182},
          doi = {10.1051/0004-6361/202348469},
archivePrefix = {arXiv},
       eprint = {2312.04914},
 primaryClass = {astro-ph.GA},
       adsurl = {https://ui.adsabs.harvard.edu/abs/2024A&A...682A.182G}
}

@ARTICLE{Murphy2011,
       author = {{Murphy}, E.~J. and {Condon}, J.~J. and {Schinnerer}, E. and {Kennicutt}, R.~C. and {Calzetti}, D. and {Armus}, L. and {Helou}, G. and {Turner}, J.~L. and {Aniano}, G. and {Beir{\~a}o}, P. and {Bolatto}, A.~D. and {Brandl}, B.~R. and {Croxall}, K.~V. and {Dale}, D.~A. and {Donovan Meyer}, J.~L. and {Draine}, B.~T. and {Engelbracht}, C. and {Hunt}, L.~K. and {Hao}, C. -N. and {Koda}, J. and {Roussel}, H. and {Skibba}, R. and {Smith}, J. -D.~T.},
        title = "{Calibrating Extinction-free Star Formation Rate Diagnostics with 33 GHz Free-free Emission in NGC 6946}",
      journal = {\apj},
         year = 2011,
        month = aug,
       volume = {737},
       number = {2},
          eid = {67},
        pages = {67},
          doi = {10.1088/0004-637X/737/2/67},
archivePrefix = {arXiv},
       eprint = {1105.4877},
 primaryClass = {astro-ph.CO},
       adsurl = {https://ui.adsabs.harvard.edu/abs/2011ApJ...737...67M}
}

@ARTICLE{Sanders2003,
       author = {{Sanders}, D.~B. and {Mazzarella}, J.~M. and {Kim}, D. -C. and {Surace}, J.~A. and {Soifer}, B.~T.},
        title = "{The IRAS Revised Bright Galaxy Sample}",
      journal = {\aj},
         year = 2003,
        month = oct,
       volume = {126},
       number = {4},
        pages = {1607-1664},
          doi = {10.1086/376841},
archivePrefix = {arXiv},
       eprint = {astro-ph/0306263},
 primaryClass = {astro-ph},
       adsurl = {https://ui.adsabs.harvard.edu/abs/2003AJ....126.1607S}
}

@ARTICLE{Norris1995,
       author = {{Norris}, R.~P. and {Forbes}, Duncan A.},
        title = "{Radio Detection of a Double Nucleus in the Merging Galaxy NGC 3256}",
      journal = {\apj},
         year = 1995,
        month = jun,
       volume = {446},
        pages = {594},
          doi = {10.1086/175818},
       adsurl = {https://ui.adsabs.harvard.edu/abs/1995ApJ...446..594N}
}

@ARTICLE{Appleton2022,
       author = {{Appleton}, P.~N. and {Emonts}, B. and {Lisenfeld}, U. and {Falgarone}, E. and {Guillard}, P. and {Boulanger}, F. and {Braine}, J. and {Ogle}, P. and {Struck}, C. and {Vollmer}, B. and {Yeager}, T.},
        title = "{The CO Emission in the Taffy Galaxies (UGC 12914/15) at 60 pc Resolution. I. The Battle for Star Formation in the Turbulent Taffy Bridge}",
      journal = {\apj},
         year = 2022,
        month = jun,
       volume = {931},
       number = {2},
          eid = {121},
        pages = {121},
          doi = {10.3847/1538-4357/ac63b2},
archivePrefix = {arXiv},
       eprint = {2203.17142},
 primaryClass = {astro-ph.GA},
       adsurl = {https://ui.adsabs.harvard.edu/abs/2022ApJ...931..121A}
}

@ARTICLE{Michiyama2020,
       author = {{Michiyama}, Tomonari and {Iono}, Daisuke and {Nakanishi}, Kouichiro and {Ueda}, Junko and {Saito}, Toshiki and {Yamashita}, Takuji and {Bolatto}, Alberto and {Yun}, Min},
        title = "{Star Formation Traced by Optical and Millimeter Hydrogen Recombination Lines and Free-Free Emissions in the Dusty Merging Galaxy NGC 3256{\textemdash}MUSE/VLT and ALMA Synergy}",
      journal = {\apj},
         year = 2020,
        month = jun,
       volume = {895},
       number = {2},
          eid = {85},
        pages = {85},
          doi = {10.3847/1538-4357/ab88a5},
archivePrefix = {arXiv},
       eprint = {2004.06123},
 primaryClass = {astro-ph.GA},
       adsurl = {https://ui.adsabs.harvard.edu/abs/2020ApJ...895...85M}
}

@ARTICLE{Komugi2012,
       author = {{Komugi}, S. and {Tateuchi}, K. and {Motohara}, K. and {Takagi}, T. and {Iono}, D. and {Kaneko}, H. and {Ueda}, J. and {Saitoh}, T.~R. and {Kato}, N. and {Konishi}, M. and {Koshida}, S. and {Morokuma}, T. and {Takahashi}, H. and {Tanab{\'e}}, T. and {Yoshii}, Y.},
        title = "{The Schmidt-Kennicutt Law of Matched-age Star-forming Regions; Pa{\ensuremath{\alpha}} Observations of the Early-phase Interacting Galaxy Taffy I}",
      journal = {\apj},
         year = 2012,
        month = oct,
       volume = {757},
       number = {2},
          eid = {138},
        pages = {138},
          doi = {10.1088/0004-637X/757/2/138},
archivePrefix = {arXiv},
       eprint = {1208.0315},
 primaryClass = {astro-ph.CO},
       adsurl = {https://ui.adsabs.harvard.edu/abs/2012ApJ...757..138K}
}

@ARTICLE{Jarrett1999,
       author = {{Jarrett}, T.~H. and {Helou}, G. and {Van Buren}, D. and {Valjavec}, E. and {Condon}, J.~J.},
        title = "{A Near- and Mid-Infrared Study of the Interacting Galaxy Pair UGC 12914/12915: ``Taffy''}",
      journal = {\aj},
         year = 1999,
        month = nov,
       volume = {118},
       number = {5},
        pages = {2132-2147},
          doi = {10.1086/301080},
archivePrefix = {arXiv},
       eprint = {astro-ph/0004268},
 primaryClass = {astro-ph},
       adsurl = {https://ui.adsabs.harvard.edu/abs/1999AJ....118.2132J}
}

@ARTICLE{Kennicutt1998,
       author = {{Kennicutt}, Jr., Robert C.},
        title = "{The Global Schmidt Law in Star-forming Galaxies}",
      journal = {\apj},
         year = 1998,
        month = may,
       volume = {498},
       number = {2},
        pages = {541-552},
          doi = {10.1086/305588},
archivePrefix = {arXiv},
       eprint = {astro-ph/9712213},
 primaryClass = {astro-ph},
       adsurl = {https://ui.adsabs.harvard.edu/abs/1998ApJ...498..541K}
}

@ARTICLE{Roussel2001,
       author = {{Roussel}, H. and {Sauvage}, M. and {Vigroux}, L. and {Bosma}, A.},
        title = "{The relationship between star formation rates and mid-infrared emission in galactic disks}",
      journal = {\aap},
         year = 2001,
        month = jun,
       volume = {372},
        pages = {427-437},
          doi = {10.1051/0004-6361:20010498},
archivePrefix = {arXiv},
       eprint = {astro-ph/0104088},
 primaryClass = {astro-ph},
       adsurl = {https://ui.adsabs.harvard.edu/abs/2001A&A...372..427R}
}

@ARTICLE{Haynes1984,
       author = {{Haynes}, M.~P. and {Giovanelli}, R.},
        title = "{Neutral hydrogen in isolated galaxies. IV. Results for the Arecibo sample.}",
      journal = {\aj},
         year = 1984,
        month = jun,
       volume = {89},
        pages = {758-800},
          doi = {10.1086/113573},
       adsurl = {https://ui.adsabs.harvard.edu/abs/1984AJ.....89..758H}
}

@ARTICLE{Ho2007,
       author = {{Ho}, Luis C. and {Keto}, Eric},
        title = "{The Mid-Infrared Fine-Structure Lines of Neon as an Indicator of Star Formation Rate in Galaxies}",
      journal = {\apj},
         year = 2007,
        month = mar,
       volume = {658},
       number = {1},
        pages = {314-318},
          doi = {10.1086/511260},
archivePrefix = {arXiv},
       eprint = {astro-ph/0611856},
 primaryClass = {astro-ph},
       adsurl = {https://ui.adsabs.harvard.edu/abs/2007ApJ...658..314H}
}

\end{document}